\documentclass[11pt]{article}
\usepackage{amsfonts,amsmath,amssymb}
\usepackage{enumerate}
\usepackage{hyperref}
\usepackage{bbm}
\usepackage{nicefrac}
\usepackage[all]{xy}
\usepackage{graphicx}
\usepackage{bm}
\usepackage{makecell}
\usepackage[table]{xcolor}

\usepackage{upgreek}
\usepackage{cancel}			
\usepackage{multicol}		
\usepackage{float}			
\usepackage{fontawesome}	
\usepackage{arydshln}		
\usepackage{subcaption}		

\usepackage{framed}

\usepackage{longtable}      

\usepackage{booktabs}

\newcommand{\be}{\begin{equation}}
\newcommand{\ee}{\end{equation}}

\newcommand{\bea}{\begin{eqnarray}}
\newcommand{\eea}{\end{eqnarray}}

\newcommand{\bes}{\begin{subequations}}
\newcommand{\ees}{\end{subequations}}

\newcommand{\cN}{{\cal N}}

\newcommand{\cZ}{{\cal Z}}

\def\sst#1{{\scriptscriptstyle #1}}

\def\0{{\sst{(0)}}}
\def\1{{\sst{(1)}}}
\def\2{{\sst{(2)}}}
\def\3{{\sst{(3)}}}
\def\4{{\sst{(4)}}}
\def\5{{\sst{(5)}}}
\def\6{{\sst{(6)}}}
\def\7{{\sst{(7)}}}
\def\8{{\sst{(8)}}}

\newcommand{\cL}{{\cal L}}

\newcommand{\vol}{\textrm{vol}}

\allowdisplaybreaks  

\usepackage{multirow}
\usepackage{rotating}

\newcommand{\ba}{\begin{align}}
\newcommand{\ea}{\end{align}}

\newcommand{\bse}{\begin{subequations}}
\newcommand{\ese}{\end{subequations}}

\allowdisplaybreaks

\renewcommand{\L}{\mathcal{L}}
\newcommand{\cy}{\mathcal{Y}}
\newcommand{\bZ}{\bar{\mathcal{Z}}}

\newcommand{\Rocket}{\rotatebox[origin=c]{270}{\faRocket}}

\begin{document}

\makeatletter
\renewcommand{\theequation}{\thesection.\arabic{equation}}
\@addtoreset{equation}{section}
\makeatother

\begin{titlepage}

\begin{flushright}

IFT-UAM/CSIC-20-103 \\

\end{flushright}

\vspace{5pt}

   \begin{center}
   \baselineskip=16pt

   \begin{LARGE}\textbf{
\mbox{A Cubic Deformation of ABJM:}}
   \end{LARGE}
   
   \vspace{10pt}

   \begin{Large}\textbf{
\mbox{The Squashed, Stretched, Warped, and Perturbed Gets Invaded}}
   \end{Large}

\vspace{25pt}

{\large  Mattia Ces\`aro$^{1}$,\ Gabriel Larios$^{1}$ \,and \,  Oscar Varela$^{1,2}$}
		
\vspace{25pt}

	\begin{small}

	{\it $^{1}$ Departamento de F\'\i sica Te\'orica and Instituto de F\'\i sica Te\'orica UAM/CSIC , \\
   Universidad Aut\'onoma de Madrid, Cantoblanco, 28049 Madrid, Spain}  \\

	\vspace{15pt}
	
	{\it $^{2}$ Department of Physics, Utah State University, Logan, UT 84322, USA}     \\	
		
	\end{small}

\vskip 50pt

\end{center}

\begin{center}
\textbf{Abstract}
\end{center}

\begin{quote}

A superpotential deformation that is cubic in one of the chiral superfields of ABJM makes the latter theory flow into a new ${\cal N}=2$ superconformal phase. This is holographically dual to a warped $\textrm{AdS}_4 \times_w S^7$ solution of M-theory equipped with a squashed and stretched metric on $S^7$. We determine the spectrum of spin-2 operators of the cubic deformation at low energies by computing the spectrum of Kaluza-Klein (KK) gravitons over the dual AdS$_4$ solution. We calculate, numerically, the complete graviton spectrum and, analytically, the spectrum of gravitons that belong to short multiplets. We also use group theory to assess the structure of the full KK spectrum, and conclude that ${\cal N}=2$ supermultiplets cannot be allocated KK level by KK level. This phenomenon, usually referred to as ``space invaders scenario", is also known to occur for another $\textrm{AdS}_4$ solution based on a different squashed $S^7$.

\end{quote}

\vfill

\end{titlepage}

\tableofcontents


\section{Introduction}


It is an interesting problem to characterise the low-energy physics that relevant deformations of superconformal field theories (SCFTs) will lead to. Focusing, for definiteness, on the Aharony-Bergman-Jafferis-Maldacena (ABJM) \cite{Aharony:2008ug} SCFT defined on a stack of M2-branes, there is a well-known relevant, mass deformation that makes the theory flow into a new infrared (IR) fixed point \cite{Benna:2008zy}. This deformation can be implemented at the level of the holomorphic superpotential by writing a new term, $({\cal Z}^4)^2$, {\it quadratic} in one of the four chiral matter superfields, ${\cal Z}^I$, $I=1, \ldots , 4$, that the $\cN=2$ formulation of ABJM encompasses. Both the resulting renormalisation group (RG) flow and its IR endpoint are manifestly $\cN=2$ and preserve an SU(3) flavour group. The IR SCFT is holographically dual to an M-theory solution written by Corrado, Pilch and Warner (CPW) \cite{Corrado:2001nv}. This geometry is a warped product, $\textrm{AdS}_4 \times_w S^7$, of four-dimensional anti-de Sitter (AdS) space and the seven-sphere, $S^7$, supported by fluxes. The metric on $S^7$ is deformed from the usual, round, SO(8)-invariant metric. Instead, it is ellipsoidally squashed and stretched along the Hopf fibre in such a way that only an $\textrm{SU}(3) \times \textrm{U(1)}$ symmetry is present, in agreement with the dual SCFT. 

Perhaps less well-known is the fact that ABJM also admits {\it cubic} deformations in the $\cN=2$ chiral superfields that are still relevant \cite{Jafferis:2011zi}. In particular, a superpotential deformation by $({\cal Z}^4)^3$ also generates an RG flow that has been similarly argued to lead the theory into a different new superconformal phase \cite{Jafferis:2011zi,Gabella:2012rc}. The dual IR geometry has been described by Gabella, Martelli, Passias and Sparks (GMPS) \cite{Gabella:2012rc} (see also \cite{Halmagyi:2012ic}). Like the CPW solution, the GMPS geometry is also an $\cN=2$ warped product $\textrm{AdS}_4 \times_w S^7$ supported by fluxes, with the metric on the internal $S^7$ deformed as well from its usual SO(8)-invariant round form by squashing and stretching the $S^7$ Hopf fibre. Accordingly, the symmetry and the supersymmetry of the GMPS configuration are also $\textrm{SU}(3) \times \textrm{U(1)}$ and $\cN=2$, again in agreement with the dual IR fixed point. Both the CPW and GMPS geometries in fact arise as particular solutions of the local analysis of \cite{Gabella:2012rc}, where the general $\cN=2$ configurations of M-theory containing an AdS$_4$ factor were classified. 

Structurally, the holomorphic superpotential deformations of ABJM by a term of the schematic form
\begin{equation} \label{eq:GenericWp}
\Delta W = ( {\cal Z}^4)^p \; , 
\end{equation}
with $p=2$ or $p=3$, are thus very similar, both from the field theory and from the gravity points of view. There is, however, a crucial difference: the existence in the former case of a related consistent truncation of $D=11$ supergravity on $S^7$ \cite{deWit:1986iy} down to maximal supergravity in four dimensions with SO(8) gauging \cite{deWit:1982ig}. The $\bm{35}_v$ scalars and $\bm{35}_c$ pseudoscalars of the $D=4$ $\cN=8$ supergravity are holographically dual to the boson and fermion mass terms of the matter superfields of $\cN=8$-enhanced ABJM, with respective relevant dimensions $\Delta=1$ and $\Delta=2$. These are precisely the type of deformations that (\ref{eq:GenericWp}) with $p=2$ induces in the field theory Lagrangian, and this is in turn the reason why this deformation is amenable to analysis within $D=4$ gauged supergravity. In contrast, the superpotential deformation (\ref{eq:GenericWp}) with $p=3$ induces interaction terms in the field theory Lagrangian among operators of dimensions $\Delta = 2$ and $\Delta = \tfrac52$ in the $\mathbf{294}_v$ and $\mathbf{224}_{cv}$ representations of SO(8), respectively. In the bulk, these operators are dual to higher Kaluza-Klein (KK) modes, and there is no known $D=4$ gauged supergravity that incorporates consistently their full non-linear interactions. 

The existence of the $\cN=8$ consistent truncation \cite{deWit:1986iy} greatly facilitates the analysis of the $p=2$ case over its $p=3$ counterpart, both from the boundary and the bulk perspectives. Firstly, the fact that the CPW solution is known analytically while the GMPS one is only known numerically can certainly be put down to the additional insight that  the consistent truncation brings in: from a purely $D=11$ perspective, both solutions are described by the same system of complicated non-linear ordinary differential equations (ODEs) in one of the internal $S^7$ angles \cite{Gabella:2012rc}. In fact, the $p=2$ AdS solution and flow from ABJM were first found in the $D=4$ gauged supergravity in \cite{Warner:1983vz} and \cite{Ahn:2000aq,Ahn:2000mf} (see also \cite{Bobev:2009ms}), respectively, and then uplifted \cite{Corrado:2001nv} to eleven dimensions. 

Secondly, as we will see in this paper, notable differences occur in the determination of the spectrum of single-trace operators with conformal dimensions of order one for both IR SCFTs. Recall that this operator spectrum can be determined holographically by classifying the KK perturbations about the dual $\textrm{AdS}_4 \times_w S^7$ solutions. For the $p=2$ CPW background, various subsectors of the KK spectrum are known. An early computation of the KK spectra for fields of all spin $ 0 \leq s \leq 2$ within the slice of KK modes contained in $D=4$ $\cN=8$ gauged supergravity was made in \cite{Nicolai:1985hs}, using a combination of supergravity and group theory methods. Using similar group theory techniques, the KK modes with spin $ 0 \leq s \leq 2$ that lie in short representations of the supersymmetry superalgebra OSp$(4|2)$ of the background were determined in \cite{Klebanov:2008vq}. The entire spectrum of KK $s=2$ gravitons was later computed in \cite{Klebanov:2009kp} (see also the more recent \cite{Dimmitt:2019qla}). Also recently, the masses for all fields with $ 0 \leq s \leq 2$ in the first KK level have been computed \cite{Malek:2019eaz}.

\newpage 

In this paper, we provide steps towards the holographic determination of the spectrum of single-trace operators of dimension of order one for the $p=3$ IR SCFT, and give complete results in specific subsectors focusing for simplicity on the spin-2 spectrum. More precisely, we determine numerically the complete spectrum of spin-2 operators. We also give analytically the spectrum of spin-2 operators that lie in short multiplets of OSp$(4|2)$. We do this by studying appropriate KK perturbations about the squashed, stretched, and warped solution of GMPS \cite{Gabella:2012rc}. We also use group theory to propose an allocation of KK modes of all spin $ 0 \leq s \leq 2$ in OSp$(4|2)$ supermultiplets. A crucial difference with respect to the analogue group theory calculations of \cite{Nicolai:1985hs,Klebanov:2008vq} for the CPW solution is that the resulting OSp$(4|2)$ representations {\it do not} descend KK level by KK level from representations of the OSp$(4|8)$ superalgebra of $\cN=8$ ABJM. Instead, the states that furnish certain supermultiplets must be drawn from different KK levels of the $\cN=8$ phase. We refer to this phenomenon as {\it space invaders scenario}, borrowing the phrase from \cite{Duff:1986hr}, where a similar phenomenon was observed for the KK spectrum on the squashed $S^7$ solution of \cite{Awada:1982pk}. The presence of space invader modes appears to be a feature of the KK spectrum of $\textrm{AdS}_4 \times S^7$ backgrounds, like GMPS \cite{Gabella:2012rc} or the squashed $S^7$ of \cite{Awada:1982pk}, that do not uplift from $D=4$ $\cN=8$ SO(8)-gauged supergravity.

Section \ref{sec: review} reviews the boundary and bulk sides of the ABJM deformation (\ref{eq:GenericWp}) with $p=3$, while section \ref{sec: spin2spec} contains our main results: the complete numerical spectrum of KK gravitons (restricted for presentation reasons up to KK level $n=3$) and some analytic results. The latter include the spectrum of gravitons that belong to short OSp$(4|2)$ supermultiplets, and a specific tower of gravitons that belong to long supermultiplets. Section \ref{sec: spaceinvaders} closes the main body of the paper with comments on the space invaders scenario. Two appendices on relevant group theory and on further speculation about space invasion complete the paper. Other results on KK graviton spectra in related contexts include \cite{Klebanov:2009kp,Bachas:2011xa,Richard:2014qsa,Pang:2015rwd,Passias:2016fkm,Pang:2017omp,Passias:2018swc,Gutperle:2018wuk,Chen:2019ydk,Speziali:2019uzn,Andriot:2019hay,Dimmitt:2019qla,Apruzzi:2019ecr}.


\section{A cubic superpotential deformation of $\cN=8$ ABJM}	\label{sec: review}

\addtocontents{toc}{\setcounter{tocdepth}{2}}


We start by reviewing some useful aspects of the field theories and their dual $\textrm{AdS}_4 \times_w S^7$ M-theory backgrounds. 

\subsection{Field theory side} \label{sec: CFT}

The ABJM theory \cite{Aharony:2008ug} is the superconformal U(N)$\times$U(N) Chern-Simons-matter gauge theory with $\mathcal{N}=6$ supersymmetry describing the worldvolume of a stack of M2-branes on a $\mathbb{C}^4/\mathbb{Z}_k$ orbifold singularity. In $\cN=2$ superfield language, its field content comprises gauge and chiral superfields. The gauge superfields $\mathcal{V}^a_b$ and $\hat{\mathcal{V}}^{\hat{a}}_{\hat{b}}$, with $a,\hat{a}$ labelling the fundamental of each U(N) factor, are governed by a Chern-Simons action at levels $k$ and $-k$ respectively. The matter superfields are $(\cZ^\mathsf{A})^a_{\hat{a}}$ and  $(\mathcal{W}_\mathsf{A})_a^{\hat{a}}$, with $\mathsf{A}=1,2$, transforming in the $({\mathbf{N}},\bar{\mathbf{N}})$ and $( \bar{\mathbf{N}}, {\mathbf{N}})$ of the gauge group as well as in the fundamental of two global SU(2)'s. Apart from the standard kinetic term for the chiral matter, the theory also contains the quartic superpotential 
\begin{equation}	\label{SuperpotSU(2)SU(2)}
	W=\frac{2\pi}{k}\epsilon_{\mathsf{A}\mathsf{C}}\epsilon^{\mathsf{B}\mathsf{D}}\, \text{tr}(\mathcal{Z}^{\mathsf{A}}\mathcal{W}_{\mathsf{B}}\mathcal{Z}^{\mathsf{C}}\mathcal{W}_{\mathsf{D}}) \;.
\end{equation}
The theory is manifestly invariant under U(1)$_R\times$SU(2)$\times$SU(2). However, for $k=1,2$, supersymmetry is expected to enhance to $\mathcal{N}=8$, with the global symmetry correspondingly upgrading to a manifest U(1)$_R\times$SU(4). To make the theory manifestly invariant under this larger group, t'Hooft monopole operators \cite{THOOFT19781}  must be used, see {\it e.g.}~\cite{Klebanov:2009sg}. These operators, $(\mathcal{M}^{q})_{\hat{a}_1,...,\hat{a}_q}^{a_1,...,a_q}$, carry $q$ units of the baryonic U(1)$_b$ flux, with U(1)$_b\subset$ U(N)$\times$U(N) being the linear combination of U(1)'s orthogonal to the one corresponding to the centre of mass of the branes. With the help of these monopole operators, a new set of chiral superfields $\mathcal{Z}^I=(\mathcal{Z}^1,\mathcal{Z}^2, \mathcal{Z}^3, \mathcal{Z}^4)$ in the fundamental of SU(4) and in the $({\mathbf{N}},\bar{\mathbf{N}})$ of the gauge group, can be introduced related to the original ABJM ones as
\begin{equation}
	(\mathcal{Z}^3)^a_{\hat{a}}=(\mathcal{W}^1)^{\hat{b}}_b(\mathcal{M}^{2})_{\hat{a}\hat{b}}^{ab}\; , \qquad 
	(\mathcal{Z}^4)^a_{\hat{a}}=(\mathcal{W}^2)^{\hat{b}}_b(\mathcal{M}^{2})_{\hat{a}\hat{b}}^{ab}\;.
\end{equation}
The SU(4)-invariant \cite{Benna:2008zy,Gustavsson:2009pm} superpotential can then be written as 
\begin{equation} \label{eq:ABJMSuperPot}
	W=\frac{4\pi}{k} \,  (\mathcal{Z}^1)^a_{\hat{a}}(\mathcal{Z}^2)^b_{\hat{b}}(\mathcal{Z}^3)^c_{\hat{c}}(\mathcal{Z}^4)^d_{\hat{d}} \, \big[ (\mathcal{M}^{-2})^{\hat{a}\hat{c}}_{bc}(\mathcal{M}^{-2})^{\hat{b}\hat{d}}_{ad}-(\mathcal{M}^{-2})^{\hat{a}\hat{d}}_{bd}(\mathcal{M}^{-2})^{\hat{b}\hat{c}}_{ac}\big]  \; .
\end{equation}
Although not manifestly, for $k=1$ the supersymmetry of the model is increased to $\cN=8$ \cite{Gustavsson:2009pm}. The supersymmetry superalgebra is therefore OSp$(4|8)$, and the R-symmetry group contained within the superalgebra is accordingly enhanced to SO(8). 

For $\cN=8$ ABJM, the superpotential (\ref{eq:ABJMSuperPot}) can be deformed by introducing an operator quadratic ($p=2$ in the notation of the introduction) in one of the chirals  \cite{Benna:2008zy}, say ${\cal Z}^4$. This deformation obviously preserves the $\textrm{SU}(3) \subset \textrm{SU}(4) \subset \textrm{SO}(8)$ flavour group that rotates the remaining ${\cal Z}^A$, $A=1,2,3$, and is manifestly $\cN=2$, with R-symmetry $\textrm{U}(1)_2 \subset \textrm{SO}(8)$. The subindex in $\textrm{U}(1)_2$ refers to $p=2$. There is a large body of literature devoted to this case, some of which was reviewed in the introduction. Here, we will be more interested in the following deformation that is instead cubic in ${\cal Z}^4$,
\begin{equation} \label{eq: deformation}
	\Delta W=\alpha (\cZ^4)^a_{\hat{a}}(\cZ^4)^b_{\hat{b}}(\cZ^4)^c_{\hat{c}}(\mathcal{M}^{-3})^{\hat{a}\hat{b}\hat{c}}_{abc}\; ,
\end{equation}
where $\alpha $ is a coupling constant. This makes equation (\ref{eq:GenericWp}) with $p=3$ more precise. Like $p=2$, the $p=3$ deformation (\ref{eq: deformation}) also preserves a flavour group $\textrm{SU}(3) \subset \textrm{SU}(4) \subset \textrm{SO}(8)$ acting on the remaining ${\cal Z}^A$, $A=1,2,3$. The deformation (\ref{eq: deformation}) is also manifestly $\cN=2$. The IR R-charges of the chirals under the associated R-symmetry group U$(1)_3$ (the subindex now referring to $p=3$) can be computed by requiring that the total superpotential, (\ref{eq:ABJMSuperPot}) plus (\ref{eq: deformation}), has R-charge two and that the free energy be extremal \cite{Jafferis:2010un}. Assuming that the monopole operators are R-neutral, the result for these U$(1)_3$ IR R-charges is \cite{Jafferis:2011zi} 
\begin{equation}	\label{eq: Rassignmentp=3}
p=3 \; : \qquad	R_1 \equiv R(\mathcal{Z}^A)=\tfrac{4}{9} \;,\ A=1,2,3\;,	\qquad 
	R_2 \equiv R(\mathcal{Z}^4)=\tfrac{2}{3}\, ,
\end{equation}  
see (\ref{eq:Rcharges}). In contrast the $p=2$ quadratic deformation leads to U$(1)_2$ IR R-charges  \cite{Klebanov:2008vq,Jafferis:2011zi}
\begin{equation}	\label{eq: Rassignmentp=2}
p=2 \; : \qquad	R_1 \equiv R(\mathcal{Z}^A)=\tfrac{1}{3} \;,\ A=1,2,3\;,	\qquad 
	R_2 \equiv R(\mathcal{Z}^4)=1 \,.
\end{equation}  
The SU(3) flavour group of both the $p=2$ and $p=3$ IR phases is the same subgroup of the SO(8) R-symmetry of the ultraviolet (UV) $\cN=8$ ABJM theory: it is, in fact, the unique $\textrm{SU}(3) \subset \textrm{SO}(8)$. However, (\ref{eq: Rassignmentp=2}) and (\ref{eq: Rassignmentp=3}) show that the U$(1)_p$ R-symmetry groups for $p=2$ and $p=3$ are different U(1) subgroups of SO(8): they are different U(1) combinations of the $\textrm{U}(1) \times \textrm{U}(1)$ that commutes with SU(3) inside SO(8): see appendix \ref{sec: branchings}. The full (super)symmetry of these IR SCFTs is thus $\textrm{OSp}(4|2)_p \times \textrm{SU}(3)$, with U$(1)_p \subset \textrm{OSp}(4|2)_p $, where we have attached a subscript $p=2$ or $p=3$ to signify that they are different (super)groups.

It is also useful to look at the deformation at the level of the Lagrangian. Using the conventional expression for the Lagrangian that derives from a superpotential (see {\it e.g.} (3.2) of \cite{Guarino:2019snw}), the effect of the deformation 
(\ref{eq: deformation}) on top of (\ref{eq:ABJMSuperPot}) is to augment the ABJM Lagrangian with the following schematic interaction terms:
\begin{equation} \label{eq:DefLag}
	\Delta \cL=\tfrac12 |\alpha|^2 \,   (Z^4)^2(\bar{Z}_4)^2  +\tfrac{1}{2}\alpha \,  \chi^4 \chi^4 Z^4  +h.c. \; ,
\end{equation}
where the contractions occur with monopole operators, which we have suppressed to avoid cluttering. Here, $Z^I$, $\chi^I$, $I=1 , \ldots , 4$, are the scalar and fermion components of the superfield $\cZ^I$. In real notation, $Z^I$ and $\chi^I$ respectively transform in the $\bm{8}_v$ and $\bm{8}_c$ of the SO(8) R-symmetry group of $\cN=8$ ABJM. Accordingly, the operators in (\ref{eq:DefLag}) are singlets under $\textrm{SU}(3) \times \textrm{U}(1)_3 \subset \textrm{SO}(8)$ that respectively branch from the $\mathbf{294}_v$ and $\mathbf{224}_{cv}$ representations of SO(8). These operators have relevant dimension $\Delta = 2$ and $\Delta = \frac52$, and thus  do indeed generate RG flow as expected. In the $p=2$ case, the ABJM Lagrangian is instead deformed with terms $Z^4 \bar{Z}_4$ and $ \chi^4 \chi^4 +h.c.$ corresponding, up to terms in the ABJM analogue of the $\cN=4 $ super-Yang-Mills Konishi operator, to mass terms for $Z^4$ and $\chi^4$. These mass terms have canonical dimension, $\Delta =1$ and $\Delta = 2$, and are $\textrm{SU}(3) \times \textrm{U}(1)_2$ singlets respectively branching from the $\bm{35}_v$ and $\bm{35}_c$ of $\textrm{SO}(8)$.

\subsection{Gravity side} \label{sec: Bkg}

The operators that deform the $\cN=8$ ABJM Lagrangian in the $p=2$ case are dual to $\textrm{SU}(3) \times \textrm{U}(1)_2$-invariant scalar and pseudoscalar KK modes that branch from the $\bm{35}_v$ and $\bm{35}_c$ representations of SO(8), respectively. Both these modes arise at KK level $n=0$ in the spectrum of the $\cN=8$ AdS$_4 \times S^7$ Freund-Rubin (FR) solution of $D=11$ supergravity, dual to $\cN=8$ ABJM:  see \cite{Duff:1986hr} for a review and table 2 of \cite{Klebanov:2008vq} for a convenient summary. As is well-known, a consistent truncation of $D=11$ supergravity on $S^7$ exists \cite{deWit:1986iy} that retains all $n=0$ KK modes and reconstructs their full non-linear interactions. The resulting $D=4$ supergravity is $\cN=8$ and has gauge group SO(8) \cite{deWit:1982ig}. In contrast, the operators in (\ref{eq:DefLag}) that trigger the $p=3$ RG flow are dual to the $\textrm{SU}(3) \times \textrm{U}(1)_3$-invariant scalar and pseudoscalar KK modes discussed above, which arise at KK levels $n=2$ and $n=1$. There is no known consistent truncation, maximally supersymmetric or otherwise, that retains these modes\footnote{Some consistent truncations are known \cite{Gauntlett:2009zw,Cassani:2011fu,Cassani:2012pj} that retain modes up the KK towers, but not the required ones. For example, the $\cN=2$ truncation of \cite{Gauntlett:2009zw} keeps SU$(4)_s$-invariant scalar and pseudoscalar modes from KK level $n=2$, dual to irrelevant operators.}.

For this reason, unlike $p=2$, the geometry dual to the $p=3$ IR SCFT must be engineered directly in $D=11$. The general class of M-theory solutions involving $\cN=2$ supersymmetry and an AdS$_4$ factor was analysed in \cite{Gabella:2012rc}. What we are referring to here as the $p=3$ GMPS geometry is a particular solution to their formalism which the authors of \cite{Gabella:2012rc} discuss in detail. The $p=2$ CPW geometry \cite{Corrado:2001nv} can also be recovered \cite{Gabella:2012rc} as a different solution in the same class. The local form of the family of geometries that encompasses both specific solutions is \cite{Gabella:2012rc}
\begin{equation}	\label{eq: vacuum}
	d\hat{s}^2_{11} = e^{2\Delta}\left(\tfrac14ds^2(\text{AdS}_4)+ds^2_7\right)	\; , \qquad 
	G_\4 =\tfrac{m}{16}\, \vol(\text{AdS}_4)+F_\4 \; ,
\end{equation}
with AdS$_4$ of radius $L=1$ and $m$ a constant. The seven-dimensional internal metric takes on the local form 
\begin{equation}	\label{eq: GMPSmetric}
	\begin{aligned}
		&ds^2_7=\frac{f\cdot\alpha}{4\sqrt{1+(1+r^2)\alpha^2}} \, ds^2 ( \mathbb{CP}_2 ) +\frac{\alpha^2}{16}\Big[dr^2+\frac{r^2f^2}{1+r^2}(d\tilde\tau+\sigma)^2	\\
		&\qquad\quad+\frac{1+r^2}{1+(1+r^2)\alpha^2}\big(d\tilde\psi+\frac{f}{1+r^2}(d\tilde\tau+\sigma)\big)^2\Big]\; ,
	\end{aligned}
\end{equation}
in terms of coordinates $r$, $\tilde{\psi}$, $\tilde{\tau}$. The line element $ds^2 ( \mathbb{CP}_2 )$ corresponds to the Fubini-Study metric on the complex projective plane, normalised so that the Ricci tensor equals six times the metric, and $\sigma$ is a local one-form potential for the K\"ahler form $J$ on $\mathbb{CP}_2$, normalised as $d \sigma = 2 \, J$. Finally, $\alpha$ and $f$ are functions of the coordinate $r$ only, the former simply a rewrite of the warp factor:
\begin{equation}	\label{eq: warping}
	e^{6\Delta} \equiv \big(\tfrac m6 \big)^2(1+r^2+\alpha^{-2}) \; .
\end{equation}
These functions are subject to the following system of non-linear differential equations:
\begin{equation}	\label{eq: GstrRelations}
	\frac{f'}{f}=-\frac12r\alpha^2\;,	\qquad\qquad
	\frac{(r\alpha'-r^2\alpha^3)f}{\sqrt{1+(1+r^2)\alpha^2}}=-3\;,
\end{equation}
where a prime denotes derivative with respect to $r$. The vectors $\partial_{\tilde{\psi}}$ and $\partial_{\tilde{\tau}}$ are Killing, and the isometry of the metric (\ref{eq: GMPSmetric}) is manifestly $\textrm{SU}(3) \times \textrm{U}(1) \times \textrm{U}(1)$. The former vector defines the local $\cN=2$ Reeb direction corresponding to the U$(1)_p$ R-symmetry, and the latter is broken by the internal four-form $F_\4$, which we will not need to specify. The internal symmetry of the full $D=11$ configuration (\ref{eq: vacuum}) is thus $\textrm{SU}(3) \times \textrm{U}(1)_p$. 

Each solution $f$ and $\alpha$ to the system of ODEs (\ref{eq: GstrRelations}) gives rise to an $\cN=2$ solution to the equations of motion of $D=11$ supergravity of the form (\ref{eq: vacuum})--(\ref{eq: warping}). The two solutions, GMPS and CPW, of interest here correspond to specific choices of $f$ and $\alpha$ subject to the boundary conditions
\begin{equation}	\label{eq: asymptotics}
	\begin{aligned}
		f&\xrightarrow[r\to0]{}\frac{3p}{p-1}\;,	\qquad\qquad	\alpha\xrightarrow[r\to0]{}wr^{-1+1/p}\;, \qquad \textrm{with $w>0$} \;,	\\
		f\xrightarrow[r\to r_0]{}&\frac{2\sqrt{1+r^2_0}}{r_0} \, (r_0-r)\;,	\qquad\qquad	\alpha\xrightarrow[r\to r_0]{}\sqrt{\frac2{r_0(r_0-r)}}\; , 
	\end{aligned}
\end{equation}
for $p=2$ or $p=3$. For these choices, the local geometry (\ref{eq: GMPSmetric}) extends globally over $S^7$. The coordinate $r$ is globally defined and ranges in $ 0 \leq r \leq r_0$ for a solution-dependent constant $r_0$. The coordinates $\tilde{\psi}$ and $\tilde{\tau}$ are only defined locally, but can be related to globally defined angles $\psi$ and $\tau$ of period $2 \pi$ via the transformation\footnote{In the notation of \cite{Gabella:2012rc}, $\tilde\psi_{\sst{\text{here}}}=\psi_{\sst{\text{there}}}$, $\tilde\tau_{\sst{\text{here}}}=\tau_{\sst{\text{there}}}$ and $\psi_{\sst{\text{here}}}={\varphi_0}_{\sst{\text{there}}}$ and $\tau_{\sst{\text{here}}}=\varphi_{\sst{\text{there}}}$ up to orientation, as one can check for the case $p=2$ combining (4.10) of \cite{Larios:2019kbw} and (3.25) of \cite{Larios:2019lxq}.}
\begin{equation}	\label{eq: globalcoords}
	\psi=\tfrac1p \, \tilde\psi \;,	\qquad\qquad \tau=\tilde\tau+\tfrac13 \big(1-\tfrac1p \big) \tilde\psi \; 
\end{equation}
for $p=2$ or $p=3$. The global coordinates $\psi$ and $\tau$ are the angles on the Hopf fibres of $S^7$ and on the $S^5$ inside $S^7$. In terms of the globally-defined angles, the $\cN=2$ Reeb vector is
\begin{equation}	\label{eq: generalReeb}
	R=\tfrac{4(p-1)}{3p}\, \partial_\tau+\tfrac4p \, \partial_\psi \equiv 4 \,  \partial_{\tilde{\psi}} \; .
\end{equation}
\begin{figure}[t] 
\centering
	\begin{subfigure}{.5\textwidth}
		\centering
		\includegraphics[width=1.0\linewidth]{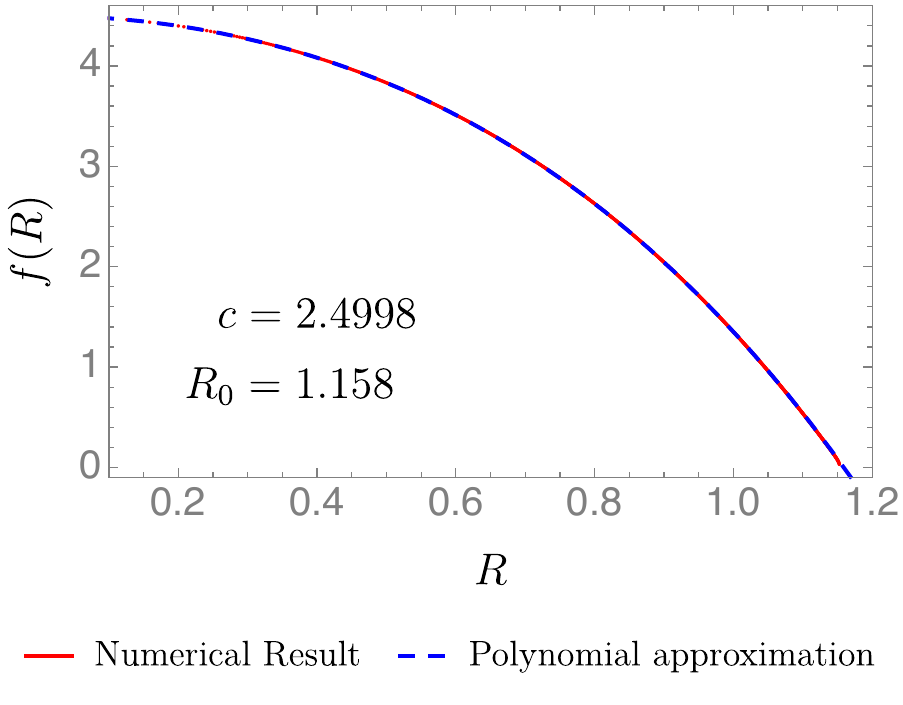}
  		\caption{}
	\end{subfigure}%
	\begin{subfigure}{.5\textwidth}
		\centering
		\includegraphics[width=1.0\linewidth]{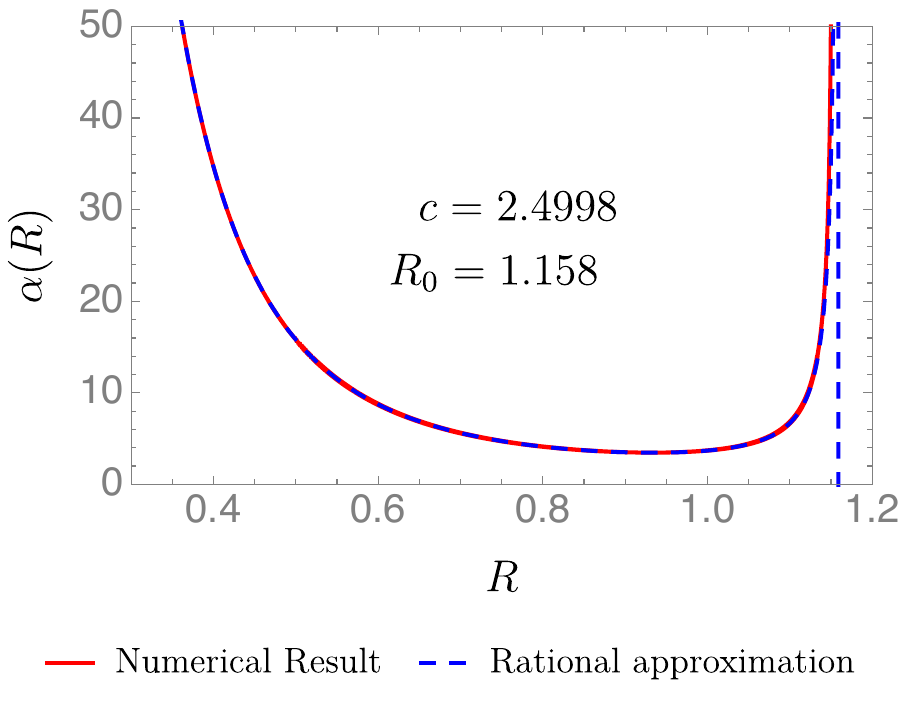}
		\caption{}
	\end{subfigure}
	\caption{Comparison between the numerically obtained functions $f$ and $\alpha$ for the GMPS solution and their respective polynomial, \eqref{eq: approxf}, and rational, \eqref{eq: approxalpha}, approximations. \label{fig:falpha} }
\end{figure}

The analytical $p=2$ CPW solution \cite{Corrado:2001nv} is recovered for \cite{Gabella:2012rc}
\begin{eqnarray} \label{eq:CPWMetricFuns}
f =6\Big(1-\frac{r}{r_0}\Big) \; , \qquad \alpha =\sqrt{\frac{2}{r(r_0-r)}} \; , \qquad r_0 =2\sqrt{2} \; .
\end{eqnarray}
The $p=3$ GMPS solution is only known numerically \cite{Gabella:2012rc}. We re-derive it here following \cite{Gabella:2012rc} in order to calibrate our numerics. The equations \eqref{eq: GstrRelations} can be combined into a single non-linear ODE for $f$,
\begin{equation} \label{eq:CombinedODE}
	\frac19f \left(R\ddot{f}-5 \dot{f}\right)+\frac{1}{3}R\dot{f}^2=\sqrt{- \dot{f} \left(6R^5 f-4 \dot{f} \left(1+R^6\right)\right)}\;,
\end{equation}
in terms of a convenient new independent variable
\begin{equation} \label{eq:R=r1/3}
R = r^{1/3} \; .
\end{equation}
In (\ref{eq:CombinedODE}), a dot denotes derivative with respect to $R$. An approximate solution to equation (\ref{eq:CombinedODE}) can be found by expanding in Taylor series about\footnote{We note a discrepancy between \eqref{eq: approxf} and (4.63) in \cite{Gabella:2012rc} in the coefficient of the $R^8$ term. We thank the anonymous referee for confirming that he agrees with \eqref{eq: approxf}.} $R=0$:
\begin{equation}	\label{eq: approxf}
	f(R)=\frac{9}{2}-c R^2-\frac{c^2}{9}R^4+\frac{\left(2187-128 c^3\right)}{3888}R^6+\frac{\left(19683 c-1264 c^4\right)}{104976}R^8+\mathcal{O}(R^{10})\;,
\end{equation}
with $c$ an integration constant. Using (\ref{eq: approxf}), the function $\alpha$ derives from \eqref{eq: GstrRelations} as 
\begin{equation}	\label{eq: approxalpha}
	\alpha^2(R)\approx\frac{4 \left(177147 R^4+26244 c \left(3 R^6-4\right)-23328 c^2 R^2-10368 c^3 R^4-5056 c^4 R^6\right)}{3 R^4 \left(1264 c^4 R^8+3456 c^3 R^6+11664 c^2 R^4-6561 c \left(3 R^6-16\right) R^2-59049 \left(R^6+8\right)\right)}\;.
\end{equation}
The approximate analytical solutions (\ref{eq: approxf}), (\ref{eq: approxalpha}) can now be used to kick off a numerical integration of the system of ODEs \eqref{eq: GstrRelations}. Imposing the right asymptotic behaviour near $R=R_0$,  given by \eqref{eq: asymptotics} with $p=3$ through (\ref{eq:R=r1/3}), the integration constant $c$ and the upper limit $r_0$ for the variable $r$ become fixed to 
\begin{equation} \label{eq:cR0}
	c\approx2.4998\;,
	\qquad\qquad
	R_0\approx1.1585 \quad \Longleftrightarrow \quad r_0\approx 1.555\;.
\end{equation}
Interestingly, the approximate solutions \eqref{eq: approxf}, \eqref{eq: approxalpha} found close to $R=0$ fit the numerically integrated functions very well across the entire range $0 \leq R \leq R_0$ for the value of $c$ in (\ref{eq:cR0}): see figure \ref{fig:falpha}.


\section{Spectrum of massive gravitons on the GMPS solution} \label{sec: spin2spec}

The spectrum of massive KK gravitons about the CPW solution \cite{Corrado:2001nv} was determined analytically in \cite{Klebanov:2009kp}. Here, we pose the analogue boundary value problem for the GMPS solution \cite{Gabella:2012rc} in section \ref{sec:BVP}, and then turn to solve it numerically in section \ref{sec:NumBVP}. The numerical integration can be systematised using the group theory of appendix \ref{sec: branchings}, and the {\it complete} graviton spectrum can be found. We do this in section \ref{sec:CompleteKK}, where we present the complete spectrum  up to KK level $n=3$. Section \ref{sec:ShortShadow} contains analytic results on the short graviton spectrum and on a specific type of long OSp$(4|2)$ supermultiplets. Finally, the analysis of the graviton spectrum is used in section \ref{sec:GMPSnotinR8} to show that the GMPS metric does not descend from the flat Euclidean metric on $\mathbb{R}^8$.

\subsection{Boundary value problem} \label{sec:BVP}

Consider the line element 
\begin{equation}	\label{eq: perturbedmetric}
	d\hat{s}^2_{11}=e^{2A}\Big[(\bar{g}_{\mu\nu}(x)+h_{\mu\nu}(x,y)\big)dx^\mu dx^\nu+d\bar{s}^2_7(y)\big]\;,
\end{equation}
where we have rescaled for convenience the warp factor and internal metric as
\begin{equation} \label{eq:WarpMet}
e^{2A}=\tfrac14 \, e^{2\Delta} \; , \qquad 
d\bar{s}^2_7 = 4 \, ds^2_7 \; ,
\end{equation}
with respect to \eqref{eq: warping} and \eqref{eq: GMPSmetric}. We fix the functions $f$ and $\alpha$ appearing in the internal squashed and stretched metric on $S^7$ and warp factor to those corresponding to the $p=3$ GMPS solution \cite{Gabella:2012rc} as reviewed in section \ref{sec: Bkg}. The external metric in (\ref{eq: perturbedmetric}) contains $\bar{g}_{\mu\nu}(x)dx^\mu dx^\nu\equiv ds^2($AdS$_4)$ as well as a spin-2 perturbation $h_{\mu\nu}(x,y)$. The latter is assumed to take on the factorised form 
\begin{equation} \label{eq:spin2Pert}
	h_{\mu\nu}(x,y) =h^{[tt]}_{\mu\nu}(x)\cy(y) \; , 
\end{equation}
with $\cy(y)$ a function on $S^7$ only, and $h^{[tt]}$ transverse ($\bar{\nabla}^\mu h^{[tt]}_{\mu\nu}$= 0) with respect to the Levi-Civita connection corresponding to $\bar{g}_{\mu\nu}$, traceless ($\bar{g}^{\mu\nu}h^{[tt]}_{\mu\nu}=0$), and subject to the Fierz-Pauli equation
\begin{equation} \label{eq:FPspin2}
	\bar{\square}h^{[tt]}_{\mu\nu}=(L^2M^2-2)h^{[tt]}_{\mu\nu} \; ,
\end{equation}
for a graviton of squared mass $M^2$. Here, $L$ is the effective AdS$_4$ radius introduced in our context by the warping $e^{2A(y)}$. The combination $L^2M^2$ is dimensionless. 

Under these assumptions, the KK graviton mass operator associated to \eqref{eq: perturbedmetric} reads \cite{Bachas:2011xa}  
\begin{equation} \label{eq:GenMassOp}
	\L=-\frac{e^{-9A}}{\sqrt{\bar{g}_{\sst{7}}}}\partial_m\big(e^{9A}\sqrt{\bar{g}_{\sst{7}}}\,\bar{g}^{mn}\partial_n\big)\,,	\qquad m,n=1,\dots,7\,,
\end{equation}
with $\bar{g}_{\sst{7}}$ and $\bar{g}^{mn}$ the determinant and inverse of the internal metric $d\bar{s}^2_7$ in (\ref{eq:WarpMet}). Using the expressions (\ref{eq: GMPSmetric}), \eqref{eq: warping}, with the former written in terms of the global coordinates \eqref{eq: globalcoords} for $p=3$, the mass operator (\ref{eq:GenMassOp}) becomes
\begin{equation} \label{eq:MassOpPart}
	\begin{aligned}
		\L=&-\frac4{r\alpha^2f^3}\partial_r\Big[rf^3\partial_r\Big] -\frac{\sqrt{1+(1+r^2)\alpha^2}}{f\cdot\alpha}\square_{S^5} 	\\
			&-\frac4{9}\Big(1+\frac1{r^2\alpha^2}\Big)\partial_{\psi}^2-\frac{8}3\Big[\frac{2}{9}\Big(1+\frac1{r^2\alpha^2}\Big)-\frac1{r^2\alpha^2f}\Big]
			\partial_{\psi}\partial_{\tau}\\
			&-\Big[-\frac{\sqrt{1+(1+r^2)\alpha^2}}{f\cdot\alpha}+\frac{16}{81}\Big(1+\frac1{r^2\alpha^2}\Big)+\frac{4(1+r^2)}{r^2\alpha^2f^2}-\frac{16}{9r^2\alpha^2f}\Big]\partial_{\tau}^2
			\; .
	\end{aligned}
\end{equation}
Here, $\square_{S^5}$ is the Laplacian on the round, unit radius $S^5$. With a graviton perturbation of the form (\ref{eq:spin2Pert}) subject to the field equation (\ref{eq:FPspin2}), the linearised Einstein equation satisfied by (\ref{eq: perturbedmetric}) becomes an eigenvalue problem for the mass operator (\ref{eq:MassOpPart}):
\begin{equation} \label{eq:spin2PDE}
\cL \, \cy = L^2M^2 \, \cy \; .
\end{equation}

At this point, we can exploit the $\textrm{SU}(3) \times \textrm{U}(1)_\tau \times \textrm{U}(1)_\psi$ isometry of the metric (\ref{eq: GMPSmetric}) and expand the $\cL$--eigenfunction $\cy$ as
\begin{equation}	\label{eq: modeexpansion}
	\cy = \sum_{\ell,m,j} \xi_{\ell,m,j}(r) \, Y_{\ell,m}(z,\bar{z}, \tau) \, e^{ij\psi} \; .
\end{equation}
Here, $\xi_{\ell,m,j}(r)$ is a function of $r$ only and $Y_{\ell,m}(z,\bar{z}, \tau)$ are the $S^5$ spherical harmonics (with definite $\textrm{U}(1)_\tau$ charge) 
\begin{equation} \label{eq:LapS5}
	\square_{S^5}Y_{\ell,m}=-\ell(\ell+4)Y_{\ell,m}\;,	\qquad\qquad
	\partial_{\tau}Y_{\ell,m}=imY_{\ell,m}\; .
\end{equation}
The quantum numbers in (\ref{eq: modeexpansion}) and (\ref{eq:LapS5}) range as
\begin{equation} \label{eq:QNranges}
\ell = 0 , 1 , 2 , \ldots \; , \qquad 
m=-\ell,-\ell+2,\dots,\ell-2,\ell \; , \qquad 
j = 0 , \, \pm 1 , \,  \pm 2  , \,  \ldots
\end{equation}
(note that $i$ in (\ref{eq: modeexpansion}) and (\ref{eq:LapS5}) is the imaginary unit). The partial differential equation (\ref{eq:spin2PDE}) thus reduces to the following Sturm-Liouville problem in $\xi_{\ell,m,j}(r)$ where, to avoid cluttering, we omit the quantum number subscripts on $\xi$: 
\begin{equation}	\label{eq: ODExi}
	\begin{aligned}
		L^2M^2\xi=& -\frac4{r\alpha^2f^3} \frac{d}{dr}\Big[rf^3  \frac{d\xi}{dr} \Big] + \frac{\sqrt{1+(1+r^2)\alpha^2}}{f\cdot\alpha}\ell(\ell+4)\xi	\\
			&+\frac4{9}\Big(1+\frac1{r^2\alpha^2}\Big)j^2\xi+\frac{8}3\Big[\frac{2}{9}\Big(1+\frac1{r^2\alpha^2}\Big)-\frac1{r^2\alpha^2f}\Big]jm\xi	\\
			&+\Big[-\frac{\sqrt{1+(1+r^2)\alpha^2}}{f\cdot\alpha}+\frac{16}{81}\Big(1+\frac1{r^2\alpha^2}\Big)+\frac{4(1+r^2)}{r^2\alpha^2f^2}-\frac{16}{9r^2\alpha^2f}\Big]m^2\xi\;.
	\end{aligned}
\end{equation}
The normalisable spin-2 modes correspond to the solutions of this ODE such that \cite{Bachas:2011xa,Richard:2014qsa}
\begin{equation}	\label{eq: normalisable}
	\int_0^{r_0} dr\, r\alpha^2 f^3\vert\xi\vert^2<\infty\; ,
\end{equation}
supplemented with the fall-offs (\ref{eq: asymptotics}) with $p=3$ for the metric functions.


\subsection{Numerics} \label{sec:NumBVP}


Solving the ODE (\ref{eq: ODExi}) on the GMPS background entails a non-trivial numerical integration over a numerical background. We have nevertheless managed to obtain the {\it complete} graviton spectrum, as we will show in section \ref{sec:CompleteKK}. In this section, we set up our numerics.

\begin{figure}[t!]
    \centering
    \begin{subfigure}[t]{0.5\textwidth}
        \centering
        \includegraphics[width=\linewidth]{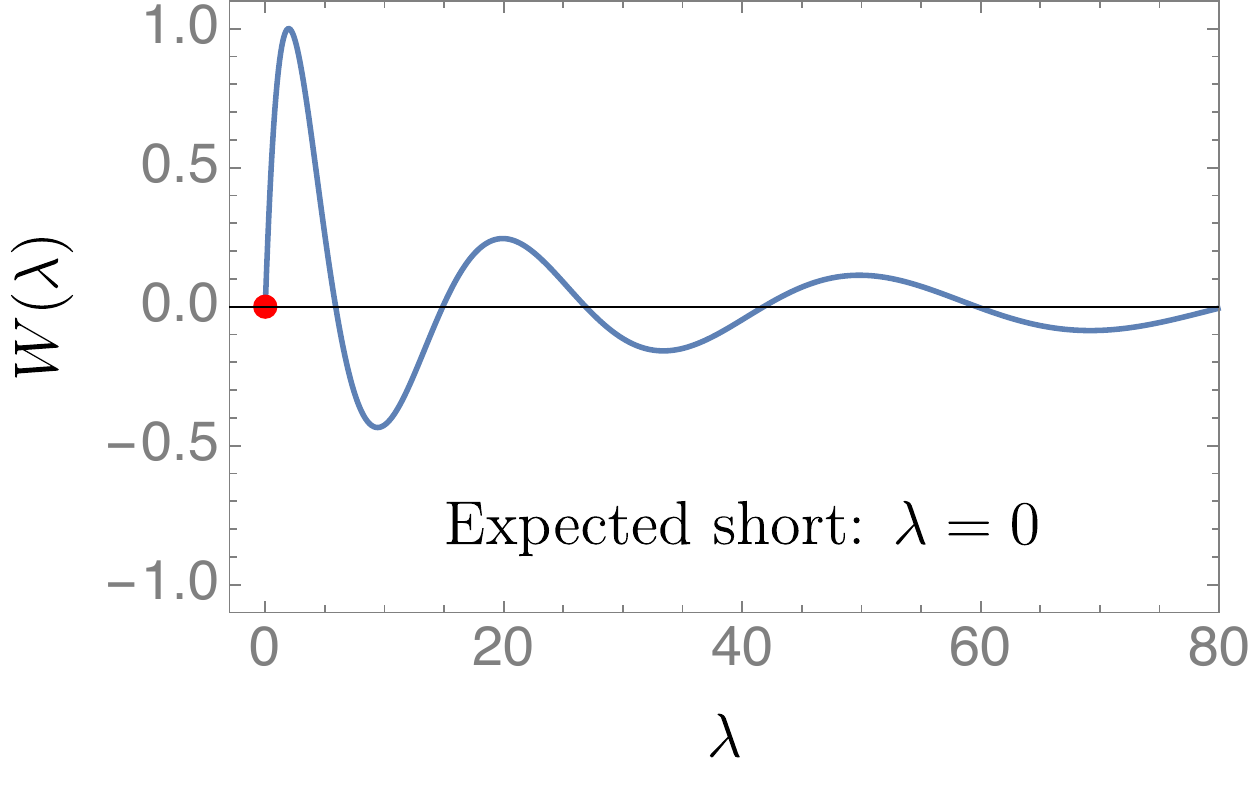}
        \caption{$j=0$}
    \end{subfigure}%
    ~ 
    \begin{subfigure}[t]{0.5\textwidth}
        \centering
        \includegraphics[width=\linewidth]{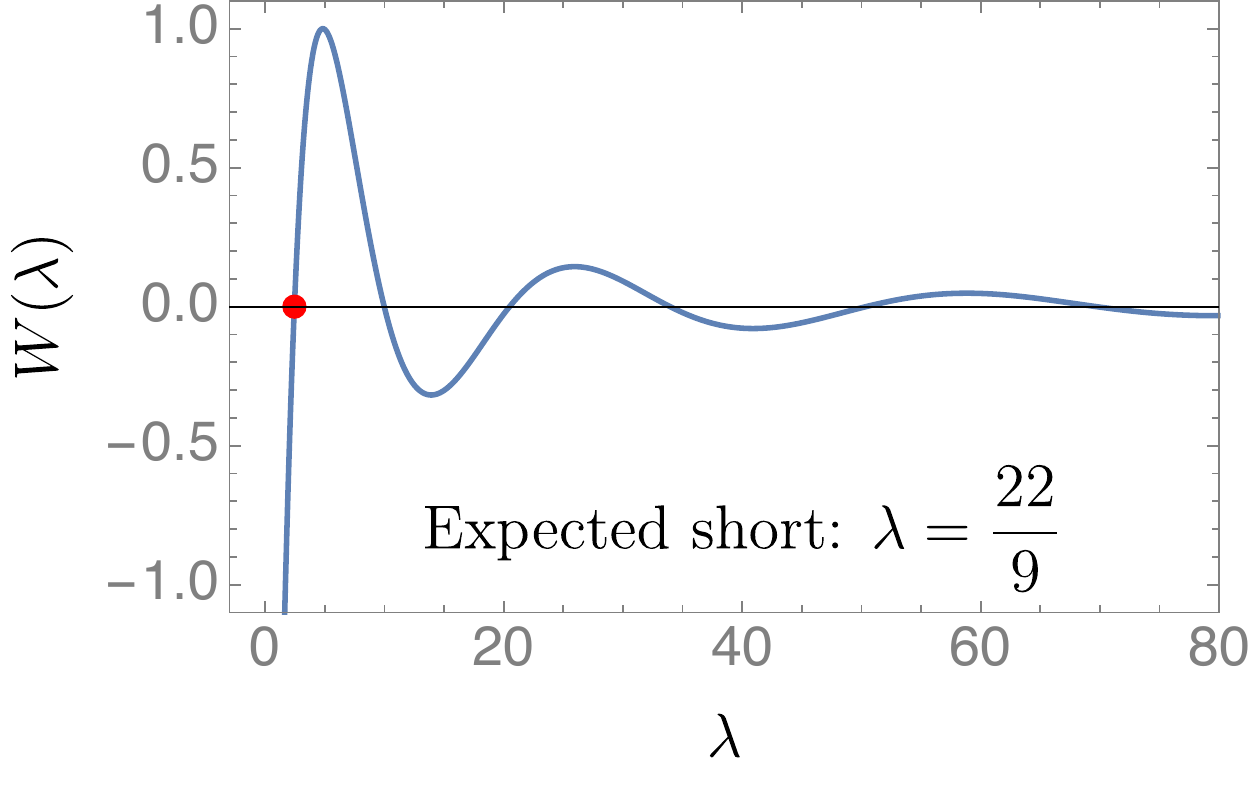}
        \caption{$j=1$}
    \end{subfigure}
    \\[5mm]
    \begin{subfigure}[t]{0.5\textwidth}
        \centering
        \includegraphics[width=\linewidth]{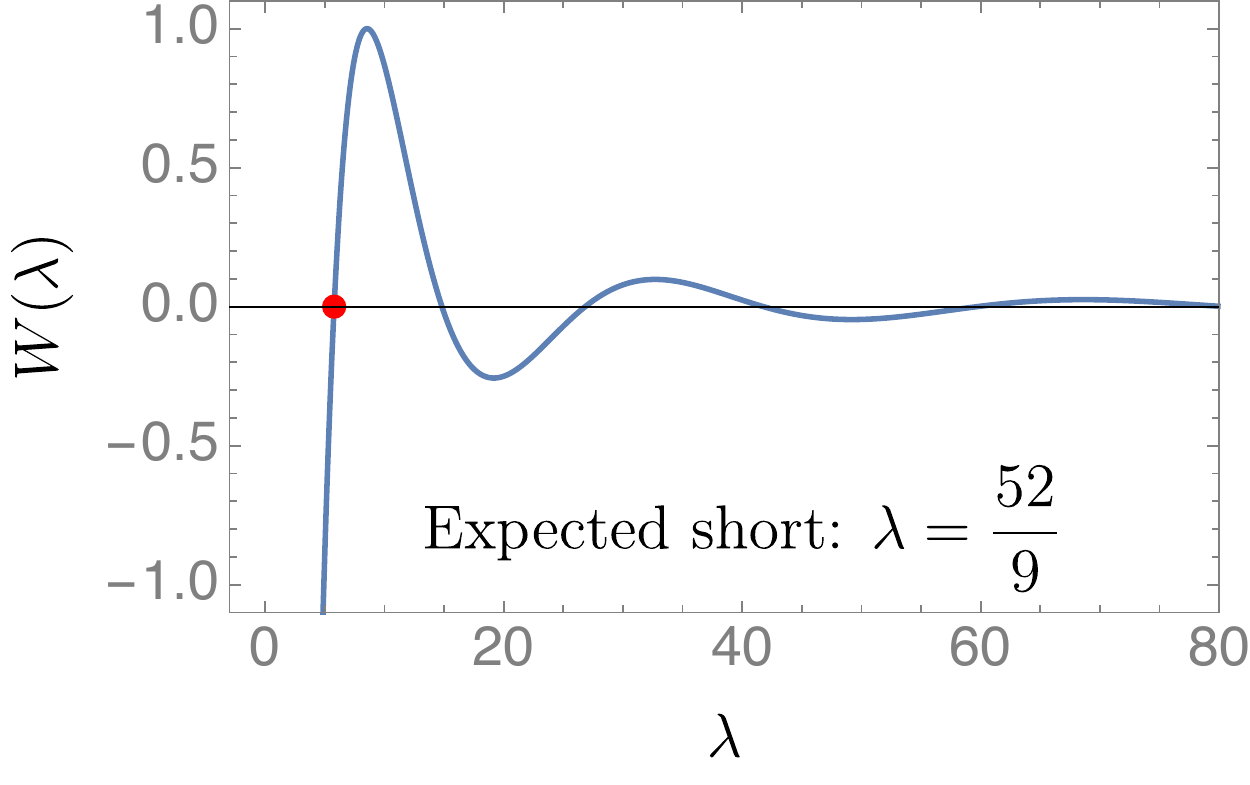}
        \caption{$j=2$}
    \end{subfigure}%
    ~ 
    \begin{subfigure}[t]{0.5\textwidth}
        \centering
        \includegraphics[width=\linewidth]{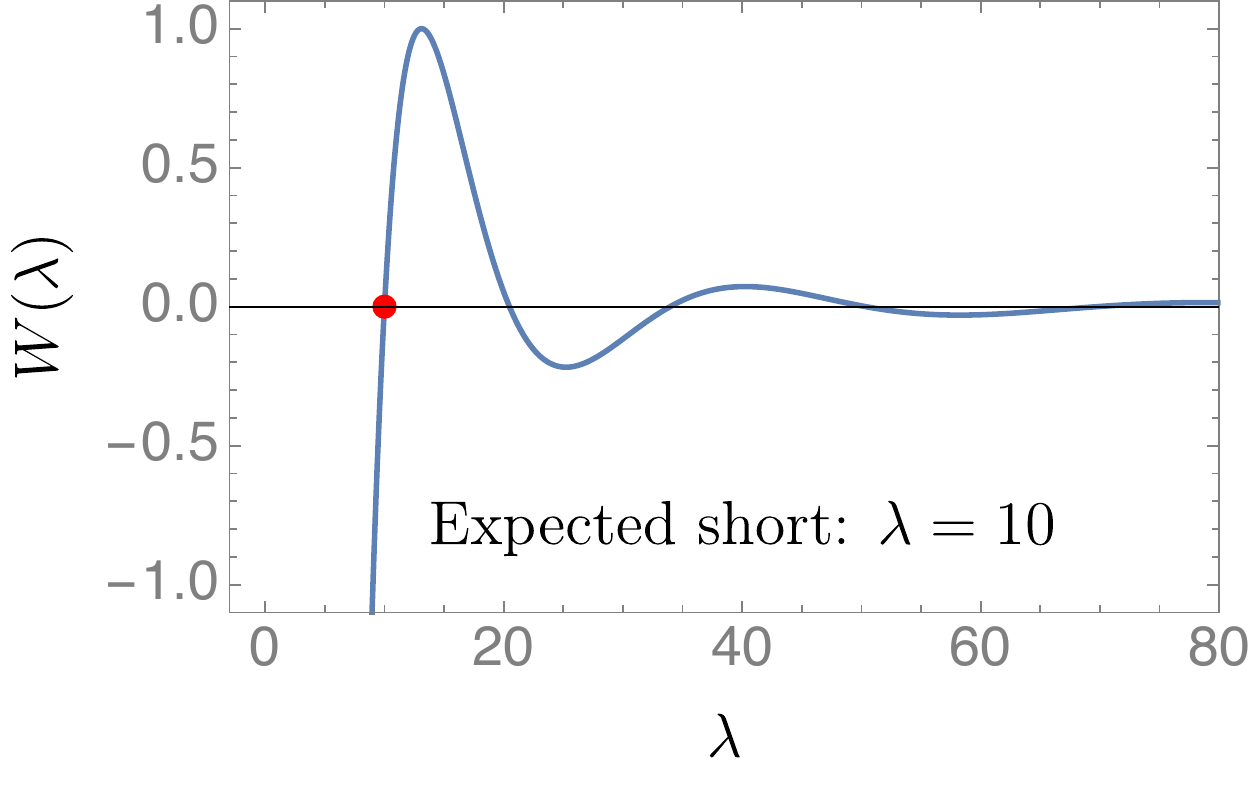}
        \caption{$j=3$}
    \end{subfigure}
    \caption{Wronskian  $W$ in (\ref{eq:Wronskian}) of the numerical functions $\xi^L_{\sst{\lambda}}(R)$ and $\xi^R_{\sst{\lambda}}(R)$ at $R=R_0/2$ for $\ell = m = 0$ and various values of $j$. The masses in table \ref{table: massesl0m0} correspond to the zeroes of $W$. The masses lying in short multiplets are marked with red dots (see section \ref{sec:ShortShadow}).}
    \label{fig: wrj03}
\end{figure}
\begin{table}[t]
	\centering
	\begin{tabular}{|c|cccc|}
		\hline
		$k\backslash\, j$	&	$0$		&	$1$		&	$2$		&	$3$		\\
		\hline
		$0$				&	0.00		&	2.44		&	5.78		&	9.99		\\
		$1$				&	5.92		&	10.00	&	14.86	&	20.54	\\
		$2$				&	14.94	&	20.57	&	26.94	&	34.11	\\
		$3$				&	27.03	&	34.13	&	42.05	&	50.71	\\
		\hline
	\end{tabular}

\caption{KK graviton masses $L^2 M^2_{k,j,\ell=0,m=0}$ on the GMPS background for a few values of the quantum numbers $k$ and $j$, at $\ell = m = 0$, as obtained from figure \ref{fig: wrj03}. The KK tower with $k=0$ corresponds to short gravitons (see section \ref{sec:ShortShadow}).}
\label{table: massesl0m0}
\end{table} 

We start by conveniently rewriting the ODE \eqref{eq: ODExi} in terms of the variable $R$ defined in (\ref{eq:R=r1/3}), whereby it becomes
\begin{equation} \label{eq:ODEinR}
	\ddot{\xi}-\Big(\frac92R^5\alpha^2-R^{-1}\Big)\dot{\xi}+\Big(\frac94 L^2M^2 R^4\alpha^2+A j^2+B\ell(\ell+4)+Cm^2+D jm\Big)\xi=0\; .
\end{equation}
Here we have defined 
\begin{eqnarray}
	A&\equiv&-\big(R^4\alpha^2+R^{-2}\big)	\;,	\nonumber\\[7pt]
	B&\equiv&-\frac94R^4\alpha f^{-1}\sqrt{1+(1+R^6)\alpha^2}	\;,	\nonumber\\[7pt]
	C&\equiv&\frac94R^4\alpha f^{-1}\sqrt{1+(1+R^6)\alpha^2}-\frac49\big(R^4\alpha^2+R^{-2}\big)-9R^{-2}(1+R^6)f^{-2}+4R^{-2}f^{-1}	\;,	\nonumber\\[7pt]
	D&\equiv&-\frac43\big(R^4\alpha^2+R^{-2}\big)+6R^{-2}f^{-1}	\;.
\end{eqnarray}
Next, we obtain asymptotic forms of the normalisable solution to (\ref{eq:ODEinR}) close to each endpoint, $R=0$ and $R=R_0$, of the domain of $R$. Near $R=0$, the asymptotic form of (\ref{eq:ODEinR}) implied by \eqref{eq: approxf} and \eqref{eq: approxalpha} depends on whether the quantum number $j$ is zero or not. For $j\neq0$, the ODE (\ref{eq:ODEinR}) close to $R=0$ takes on the form
\begin{equation} \label{eq:ODEinRjneq0}
	\ddot{\xi}+\frac{1}{R} \dot{\xi}-\frac{j^2}{R^2}\xi=0\; ,
\end{equation}
where the term in the eigenvalue $L^2M^2$ drops out as it is subleading. The ODE (\ref{eq:ODEinRjneq0})
has solutions
\begin{equation}	\label{eq: near0jneq0}
	\xi=aR^j+bR^{-j}\;,
\end{equation}
with $a,b$ constants. Compatibility with the normalisability condition \eqref{eq: normalisable} requires $a=0$ for $j<0$ and $b=0$ for $j>0$. When $j=0$, (\ref{eq:ODEinR}) close to $R=0$ reduces instead to
\begin{equation} \label{eq: near0j=0}
	\ddot{\xi}+\frac{1}{R} \dot{\xi}+\Big(\frac{2c}{3}L^2M^2-\frac{4c}{27}\ell(\ell+4)+\frac{4c}{243}m^2\Big)\xi=0\;,
\end{equation}
with the constant $c$ given in \eqref{eq:cR0}. The solutions of (\ref{eq: near0j=0}) are now
\begin{equation}	\label{eq:near0j0}
	\xi=aJ_0\Big(\sqrt{\tfrac{2c}{3}L^2M^2-\tfrac{4c}{27}\ell(\ell+4)+\tfrac{4c}{243}m^2}R\Big)
	+bY_0\Big(\sqrt{\tfrac{2c}{3}L^2M^2-\tfrac{4c}{27}\ell(\ell+4)+\tfrac{4c}{243}m^2}R\Big)\;,
\end{equation}
with $a,b$ again integration constants and $J_0$ and $Y_0$ Bessel functions. In this case, normalisability, \eqref{eq: normalisable}, requires $b=0$. 

\begin{figure}[t]
    \centering
    \begin{subfigure}[t]{0.5\textwidth}
        \centering
        \includegraphics[width=\linewidth]{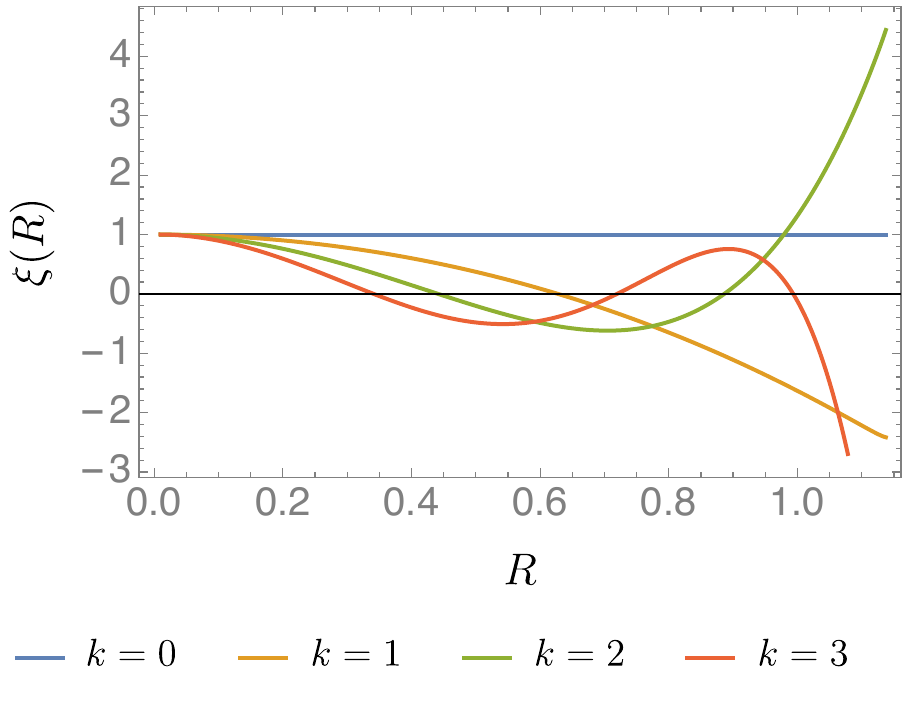}
        \caption{$j=0$}
    \end{subfigure}%
    ~ 
    \begin{subfigure}[t]{0.5\textwidth}
        \centering
        \includegraphics[width=\linewidth]{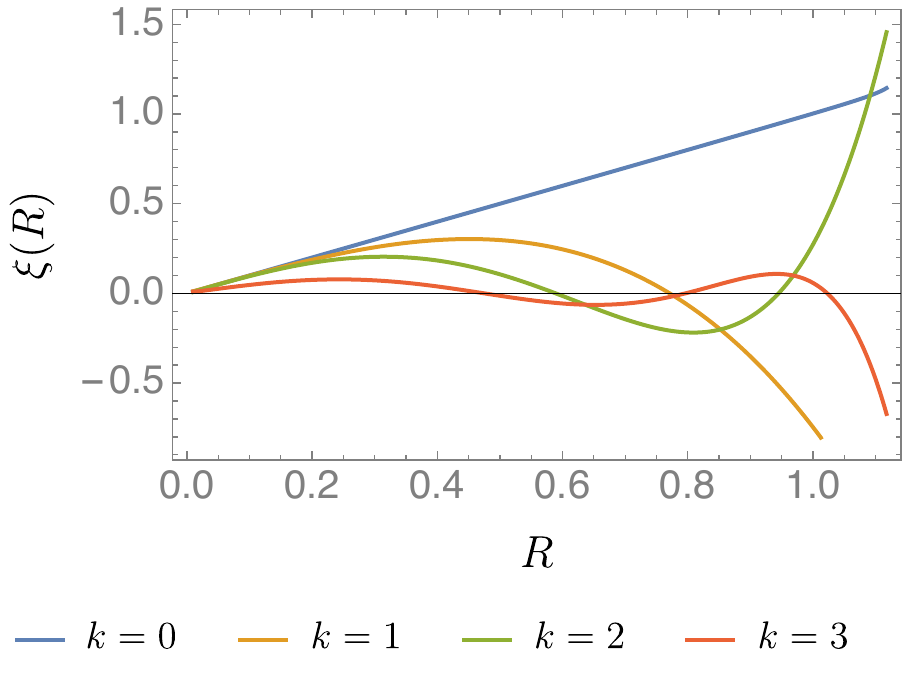}
        \caption{$j=1$}
    \end{subfigure}
    \\[5mm]
    \begin{subfigure}[t]{0.5\textwidth}
        \centering
        \includegraphics[width=\linewidth]{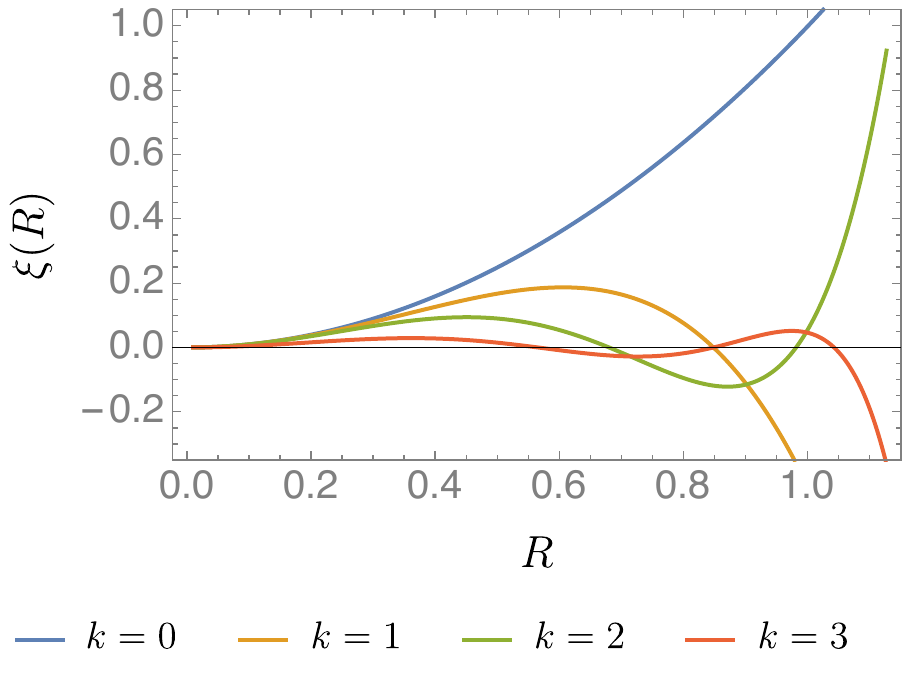}
        \caption{$j=2$}
    \end{subfigure}%
    ~ 
    \begin{subfigure}[t]{0.5\textwidth}
        \centering
        \includegraphics[width=\linewidth]{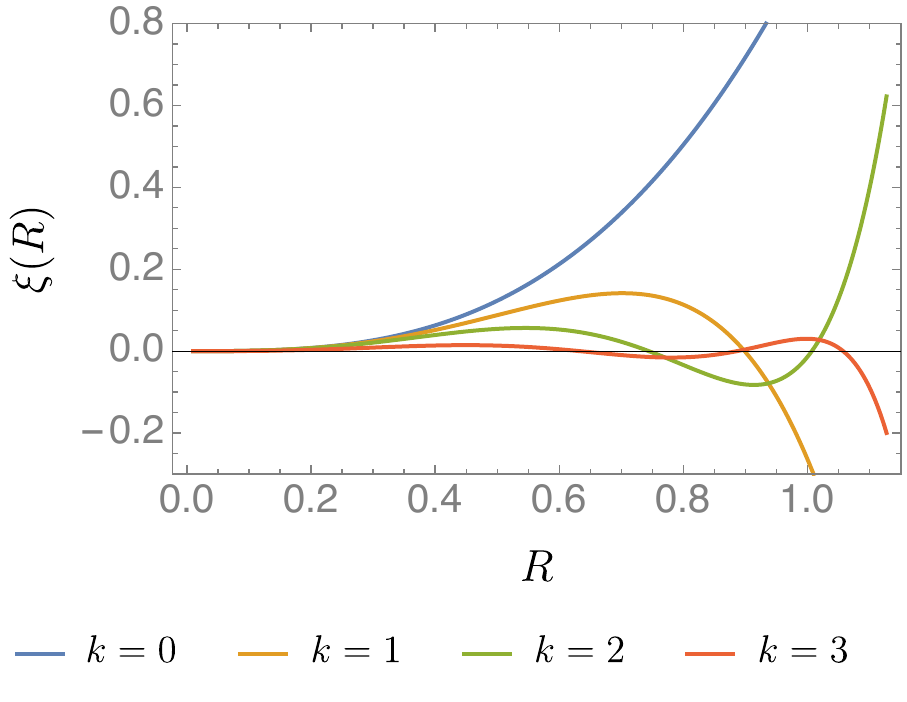}
        \caption{$j=3$}
    \end{subfigure}
    \caption{Numerical eigenfunctions for the modes with masses in table~\ref{table: massesl0m0} . }
    \label{fig: combwavefunctions}
\end{figure}

Near $R=R_0$, with $R_0$ specified in (\ref{eq:cR0}), the asymptotic form turns out to depend on the quantum number $\ell$. For $\ell=0$, (\ref{eq:ODEinR}) close to $R=R_0$ becomes
\begin{equation}
	\ddot{\xi}-\frac{3}{R_0-R}\dot{\xi}+\frac{1}{R_0(R_0-R)} \left(\frac{3}{2}L^2M^2-\frac{2}{3}j^2\right)\xi=0\; .
\end{equation}
This has solutions 
{\small
\begin{equation}\label{eq:nearR0l0}
	\xi= \frac{u}{R_0-R} I_2\left(\sqrt{\frac{2}{3}\frac{\left(4 j^2-9 L^2M^2\right) (R_0-R)}{R_0}}\right)
	+\frac{v}{R_0-R} K_2\left(\sqrt{\frac{2}{3}\frac{\left(4 j^2-9 L^2M^2\right) (R_0-R)}{R_0}}\right)\,,
\end{equation}}%
where $u,v$ are constants and $I_2$ and $K_2$ modified Bessel functions. If $ \ell \neq0$, then (\ref{eq:ODEinR}) close to $R=R_0$ can be approximated to 
\begin{equation}
	\ddot{\xi}-\frac{3}{R_0-R}\dot{\xi}-\frac{1}{4(R_0-R)^2}\ell(\ell+4)\xi=0\; ,
\end{equation}
which has solutions
\begin{equation}	\label{eq:nearR0lneq0}
	\xi=u(R_0-R)^{\ell/2}+v(R_0-R)^{-(\ell+4)/2}\;.
\end{equation}
In this case, normalisability requires $v=0$ in both \eqref{eq:nearR0l0} and \eqref{eq:nearR0lneq0}.

Now, the above asymptotic functions near $R=0$ and $R=R_0$ can be used as seeds for the numerical integration of the ODE (\ref{eq:ODEinR}). Following \cite{Richard:2014qsa}, we have performed the integration starting from both ends of the $R$ interval, in terms of a parameter $\lambda$ that labels the possible dimensionless squared masses. Denoting the functions obtained, for each $\lambda$, by integrating from the left and from the right as $\xi^L_{\sst{\lambda}}(R)$ and $\xi^R_{\sst{\lambda}}(R)$, the valid solutions to (\ref{eq:ODEinR}) can only arise for the specific values of $\lambda$ for which both $\xi^L_{\sst{\lambda}}(R)$ and $\xi^R_{\sst{\lambda}}(R)$ are linearly dependent. This requires that the Wronskian,
\begin{equation} \label{eq:Wronskian}
	W(\lambda, R)=\xi^L_{\sst{\lambda}} (R) \, \dot{\xi}^R_{\sst{\lambda}} (R) -\xi^R_{\sst{\lambda}} (R) \,  \dot{\xi} ^L_{\sst{\lambda}} (R) \; ,
\end{equation}
vanishes for all $R$ in its range. We choose, without loss of generality, to evaluate (\ref{eq:Wronskian}) at the midpoint of the interval in order to minimise the accumulated numerical error of each solution, $\xi^L_{\sst{\lambda}}(R)$ and $\xi^R_{\sst{\lambda}}(R)$. Plotting $W(\lambda, \frac{R_0}{2})$ as a function of $\lambda$ at fixed value of the quantum numbers $j$, $\ell$ and $m$, the physical masses occur at the zeros of this function: see for example figure \ref{fig: wrj03} for the $\ell = m = 0$ case. The zeroes turn out to form an infinite discrete set, which we label by a non-negative integer $k = 0, 1 , 2, \ldots$ (the first zero corresponding to $k=0$). We have tabulated a few results in table \ref{table: massesl0m0}. Finally, the eigenfunctions can be plotted numerically: see figure \ref{fig: combwavefunctions} for a few examples.


\subsection{The complete numerical KK graviton spectrum} \label{sec:CompleteKK}


Repeating the process outlined in section \ref{sec:NumBVP} for other values of the quantum numbers $j$, $\ell$ and $m$, we find other discrete series of graviton masses in the KK spectrum, labelled by a non-negative integer $k$. This procedure can be systematised using group theory by exploiting the fact that the GMPS geometry extends globally over $S^7$. In this way, we are able to find the complete KK graviton spectrum.

For the $\cN=8$ FR solution AdS$_4 \times S^7$, with $S^7$ equipped with the round metric, the gravitons at KK level $n = 0 , 1 , 2 ,  \ldots$ belong to the symmetric traceless $[n,0,0,0]$ representation of SO(8). For the squashed, stretched and warped GMPS solution AdS$_4 \times_w S^7$, the gravitons must instead arrange themselves in $\textrm{SU}(3) \times \textrm{U} (1)_3$ representations $[p,q]_{r_3}$, with $r_3$ labelling the U$(1)_3$ R-charge. Since GMPS arises as an IR fixed point of an RG flow that originates upon relevant deformation of the $\cN=8$ phase, the quantum numbers $p$, $q$ and $r_3$ must be related to the SO(8) KK level $n$ by branching of $[n,0,0,0]$ under 
\begin{equation} \label{eq:SO8toSU3U1}
\textrm{SO}(8) \supset \textrm{SU}(3) \times \textrm{U} (1)_3 \; .
\end{equation}
Specifically, we find (see appendix \ref{sec: branchings})
\begin{equation} \label{eq:branchingSymTraceless}
[n,0,0,0] \xrightarrow{{\scriptsize \text{SU}(3)\times\text{U}(1)_{3}}} \bigoplus_{\ell=0}^{n}\bigoplus_{t=0}^{n-\ell}\bigoplus_{p=0}^{\ell}
				[p,\;\ell-p]_{-R_1(\ell-2p)+R_2(n-\ell-2t)}\;,%
\end{equation}
where $R_1$ and $R_2$ are the IR R-charges (\ref{eq: Rassignmentp=3}) (or (\ref{eq: Rassignmentp=2}) for CPW) of the coordinates transverse to the M2-branes. Group theory arguments also allow us to translate between the set of quantum numbers $(k,j,\ell,m)$ used in section \ref{sec:NumBVP}, with the quantum numbers $(n, \ell , p , t)$ adapted to the branching (\ref{eq:branchingSymTraceless}):
\begin{equation} \label{eq:QNredef}
n=2k+\vert j\vert 
+ \ell \; , \qquad
m = 2p - \ell \; , \qquad 
j = n-\ell-2t  \, , 
\end{equation}
with $\ell$ here and in section \ref{sec:NumBVP} identified. Finally, it can be checked that the quantum numbers $(n, \ell , p , t)$ that characterise the KK graviton spectrum range as
\begin{equation} \label{ref:QNRanges}
n = 0 , 1 , \ldots , \qquad
\ell = 0 , 1 , \ldots , n \; , \qquad 
t = 0 , 1 , \ldots , n - \ell \; , \qquad 
p = 0 , 1 , \ldots , \ell \; , \qquad 
\end{equation}
in agreement with the branching (\ref{eq:branchingSymTraceless}). 

Integrating numerically the ODE (\ref{eq:ODEinR}) as explained in section \ref{sec:NumBVP}, but now systematically using the quantum numbers (\ref{ref:QNRanges}), we are guaranteed to sweep over the complete mass spectrum. The eigenfunctions, and thus the schematic form of the dual operators, can be similarly inferred from the branching (\ref{eq:branchingSymTraceless}). Table \ref{tab:KKGravMassSU3U1N=2SO8} summarises our results up to SO(8) KK level $n=3$.

\begin{table}[t]
\centering

\resizebox{\textwidth}{!}{

\begin{tabular}{|c|l|c|c|c|c|c|}
\hline
$n$                 & $[p,\ell-p]_{\frac49 (2p-\ell) + \frac23(n-\ell-2t) }$          & $d_{p,\,\ell-p}$ & $L^2 M^2_{n,\ell,t,p} $        & $\Delta_{n,\ell,t,p}$		& Dual operator     & Short?   \\ \hline\hline
0                   & $[0,0]_0$		& 1   & 0		&	3	&		$\mathcal{T}^{(0)}_{\alpha\beta}\vert_{s=2}$	& $\checkmark$     \\ \hline
\multirow{2}{*}{1}  & $[0,0]_{\pm \frac23}$		& 1   & $\frac{22}9$	& $\frac{11}3$	& $\mathcal{T}^{(0)}_{\alpha\beta}\mathcal{Z}^4\vert_{s=2}$, c.c. &$\checkmark$       \\ \cline{2-7}
                    & $[1,0]_{\frac{4}{9}},[0,1]_{-\frac{4}{9}}$   & 3   &	1.76	&	3.50		& $\mathcal{T}^{(0)}_{\alpha\beta}\mathcal{Z}^a\vert_{s=2}$, c.c. &        \\ \hline
\multirow{6}{*}{2}  & $[0,0]_{\pm \frac43}$		& 1   & $\frac{52}9$	& $\frac{13}3$	& $\mathcal{T}^{(0)}_{\alpha\beta}(\mathcal{Z}^4)^2\vert_{s=2}$, c.c.    &$\checkmark$  \\ \cline{2-7}
                    & $[1,0]_{-\frac{2}{9}},[0,1]_{\frac{2}{9}}$   & 3   & 4.68	&	4.13		& $\mathcal{T}^{(0)}_{\alpha\beta}\mathcal{Z}^a\bar{\mathcal{Z}}_4\vert_{s=2}$, c.c. &        \\ \cline{2-7}
                    & $[2,0]_{\frac{8}{9}},[0,2]_{-\frac{8}{9}}$   & 6   & 3.88	&	3.97		&  $\mathcal{T}^{(0)}_{\alpha\beta}\mathcal{Z}^{(a}\mathcal{Z}^{b)}\vert_{s=2}$, c.c. &       \\ \cline{2-7}
                    & $[1,0]_{\frac{10}{9}},[0,1]_{-\frac{10}{9}}$   & 3   & 5.07	&	4.21	& $\mathcal{T}^{(0)}_{\alpha\beta}\mathcal{Z}^a {\mathcal{Z}}^4\vert_{s=2}$ , c.c. &        \\ \cline{2-7}
                    & $[0,0]_0$	& 1   & 5.92	&	4.36	&	$\mathcal{T}^{(0)}_{\alpha\beta}(1-4a^2\mathcal{Z}^4\bar{\mathcal{Z}}_4+b\mathcal{Z}^a\bar{\mathcal{Z}}_a)\vert_{s=2}$  &      \\ \cline{2-7}
                    & $[1,1]_0$		& 8   & 4	&	4	&  $\mathcal{T}^{(0)}_{\alpha\beta}(\mathcal{Z}^a \bar{{\mathcal{Z}}_b}-\frac{1}{3}\delta^{a}_{b}\mathcal{Z}^c\bar{\mathcal{Z}}_c)\vert_{s=2}$&     \\ \hline
\multirow{10}{*}{3} & $[0,0]_{\pm 2}$		& 1   &	10	&	5	& $\mathcal{T}^{(0)}_{\alpha\beta}(\mathcal{Z}^4)^3\vert_{s=2}$, c.c. & $\checkmark$  \\ \cline{2-7}
                    & $[1,0]_{-\frac{8}{9}},[0,1]_{\frac{8}{9}}$   & 3   &	8.48	&	4.77		& $\mathcal{T}^{(0)}_{\alpha\beta}\mathcal{Z}^a(\bar{\mathcal{Z}}_4)^2\vert_{s=2}$, c.c &	\\ \cline{2-7}
                    & $[2,0]_{\frac{2}{9}},[0,2]_{-\frac{2}{9}}$   & 6   &	7.27	&	4.59		& $\mathcal{T}^{(0)}_{\alpha\beta}\mathcal{Z}^{(a}\mathcal{Z}^{b)}(\bar{\mathcal{Z}}_4)\vert_{s=2}$, c.c. &                 \\ \cline{2-7}
                    & $[3,0]_{\frac43},[0,3]_{-\frac43}$  	& 10  &	6.36	&	4.43	 	& $\mathcal{T}^{(0)}_{\alpha\beta}\mathcal{Z}^{(a}\mathcal{Z}^b\mathcal{Z}^{c)}\vert_{s=2}$, c.c. &       \\ \cline{2-7}
                    & $[0,0]_{\pm \frac23}$                              & 1   &	10.00	&	5.00		& $\mathcal{T}^{(0)}_{\alpha\beta}(2-5a^2\mathcal{Z}^4\bar{\mathcal{Z}}_4+b\mathcal{Z}^c\bar{\mathcal{Z}}_c)\mathcal{Z}^4\vert_{s=2}$, c.c. &      \\ \cline{2-7}
                    & $[1,0]_{\frac{16}{9}},[0,1]_{-\frac{16}{6}}$   & 3   & 9.28	&	4.90	& $\mathcal{T}^{(0)}_{\alpha\beta}\mathcal{Z}^a(\mathcal{Z}^4)^2\vert_{s=2}$, c.c. &                       \\ \cline{2-7}
                    & $[1,0]_{\frac{4}{9}},[0,1]_{-\frac{4}{9}}$   & 3   &	9.08	&	4.87		& $\mathcal{T}^{(0)}_{\alpha\beta}\mathcal{Z}^a(1-5a^2\mathcal{Z}^4\bar{\mathcal{Z}}_4+b\mathcal{Z}^c\bar{\mathcal{Z}}_c)\vert_{s=2}$, c.c. &                     \\ \cline{2-7}
                    & $[1,1]_{\pm \frac23}$                              & 8   &	$\frac{70}9$	& $\frac{14}3$	&$\mathcal{T}^{(0)}_{\alpha\beta}(\mathcal{Z}^a \bar{{\mathcal{Z}}_b}-\frac{1}{3}\delta^{a}_{b}\mathcal{Z}^c\bar{\mathcal{Z}}_c)\mathcal{Z}^4\vert_{s=2}$, c.c. &         \\ \cline{2-7}
                    & $[2,0]_{\frac{14}{9}}, [0,2]_{-\frac{14}{9}} $ & 6   & 8.02	&	4.70	&  $\mathcal{T}^{(0)}_{\alpha\beta}\mathcal{Z}^{(a}\mathcal{Z}^{b)}{\mathcal{Z}}^4\vert_{s=2}$, c.c&                      \\ \cline{2-7}
                    & $[2,1]_{\frac{4}{9}}, [1,2]_{-\frac{4}{9}} $ & 15  &	6.60	&	4.48	& $\mathcal{T}^{(0)}_{\alpha\beta}(\mathcal{Z}^{(a}\mathcal{Z}^{b)}\bar{\mathcal{Z}}_c-\delta^{(a}_{c}\mathcal{Z}^{b)}\mathcal{Z}^d\bar{\mathcal{Z}}_d)\vert_{s=2}$, c.c. &                      \\ \hline
\end{tabular}

}
\caption{ The complete KK graviton spectrum on the GMPS solution up to KK level $n=3$. For each state, the $\textrm{SU}(3) \times \textrm{U}(1)_3$ representation where it belongs is shown, along with its degeneracy $d_{p,\,\ell-p}$, mass  $L^2 M^2_{n,\ell,t,p} $,  and conformal dimension $\Delta_{n,\ell,t,p}$. The schematic form of the dual  operator is shown, with $\mathcal{T}^{(0)}_{\alpha\beta}$ denoting the IR SCFT stress-energy operator. Masses that correspond to short multiplets (ticked in the last column) and {\it shadow} long multiplets have been given analytically: see section \ref{sec:ShortShadow}. 
}
\label{tab:KKGravMassSU3U1N=2SO8}
\end{table}

\newpage

\subsection{Analytic results: short and shadow gravitons}  \label{sec:ShortShadow}

In the previous section, we arranged the GMPS graviton spectrum in representations of the $\textrm{SU}(3) \times \textrm{U} (1)_3$ residual bosonic symmetry of the background. This geometry also preserves $\cN=2$ supersymmetry, so the graviton spectrum must organise itself into representations of the full (super)symmetry group OSp$(4|2) \times \textrm{SU}(3)$ (with the $ \textrm{U} (1)_3$ R-symmetry contained in the  OSp$(4|2)$ factor). Recall that there are three types of OSp$(4|2)$ multiplets that contain states up to spin $s=2$: massless, short and long. See {\it e.g.} tables 8, 9 and 10 of \cite{Klebanov:2008vq} for a summary of their state contents. 

From table \ref{tab:KKGravMassSU3U1N=2SO8} we see that we obtain, as expected, a massless graviton which is an $\textrm{SU}(3) \times \textrm{U} (1)_3$ singlet. In addition to the $D=4$ metric and gravitini, the $\cN=2$ massless graviton multiplet contains a vector. A fully non-linear consistent truncation on GMPS \cite{Larios:2019lxq} (and on CPW \cite{Larios:2019kbw,Larios:2019lxq}) beyond the linearised analysis presented here exists to this $D=4$ field content. This is in agreement with the general statements of \cite{Gauntlett:2007ma,Cassani:2019vcl}. 

Inspection of our numerical results also allows us to detect analytically a tower of short gravitons. We indeed observe that, for every $n$, our numerical eigenvalues for the states with $\textrm{SU}(3) \times \textrm{U} (1)_3$ quantum numbers $[0,0]_{\pm R_2 n }$, with $R_2$ given by the R-charge of ${\cal Z}^4 $ in (\ref{eq: Rassignmentp=3}), are very well approximated by the analytic expression
\begin{equation}	\label{eq: massshorts}
	L^2M^2_{n}= R_2 n \,  \big(R_2  n +3 \big)\; .
\end{equation}
These states are thus short, since their conformal dimensions
\begin{equation} \label{eq:ConfDimShort}
\Delta_n = R_2  n +3 \; , 
\end{equation}
which arise from (\ref{eq: massshorts}) as the larger solution to the  equation 
\begin{equation} \label{eq:DeltaM2}
\Delta (\Delta -3 ) = M^2L^2 \; ,
\end{equation}
are locked in terms of their R-symmetry charges
\begin{equation} \label{eq:RchargeShort}
R_n = \pm R_2 n
\end{equation}
through the relation
\begin{equation} \label{eq:ShortDelta}
\Delta_n = |R_n | +3 \; .
\end{equation}
For these states, the numerically obtained value of the masses has been replaced in table \ref{tab:KKGravMassSU3U1N=2SO8} with the analytic value (\ref{eq: massshorts}). 

\begin{figure}[t]
\centering
	\begin{subfigure}[t]{.33\textwidth}
		\centering
		\includegraphics[width=1.0\linewidth]{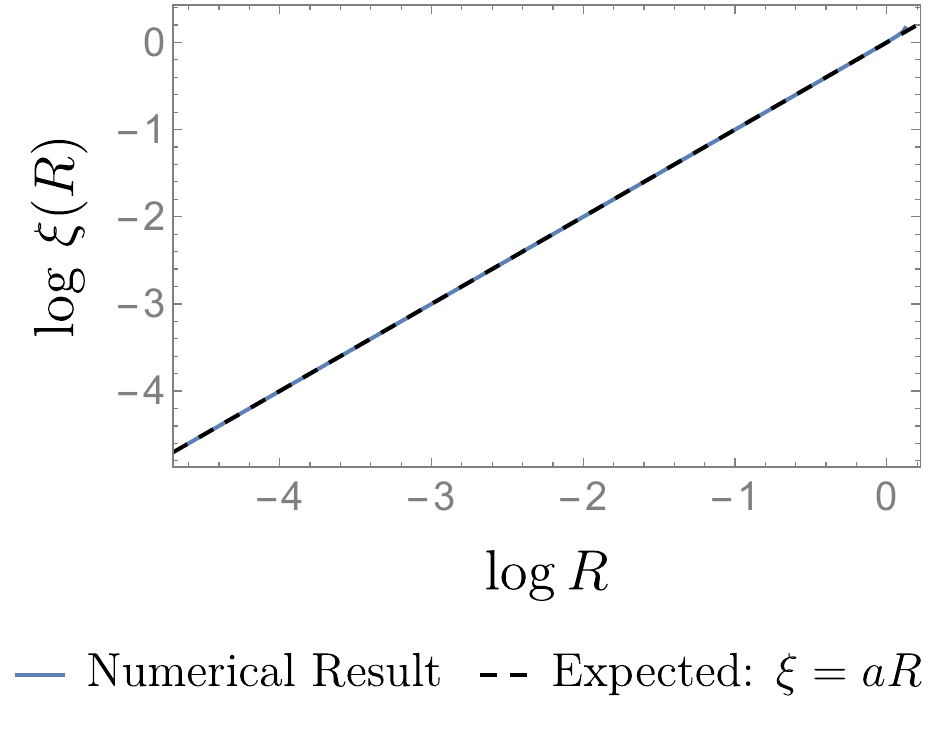}
  		\caption*{}
	\end{subfigure}~
	\begin{subfigure}[t]{.33\textwidth}
		\centering
		\includegraphics[width=1.0\linewidth]{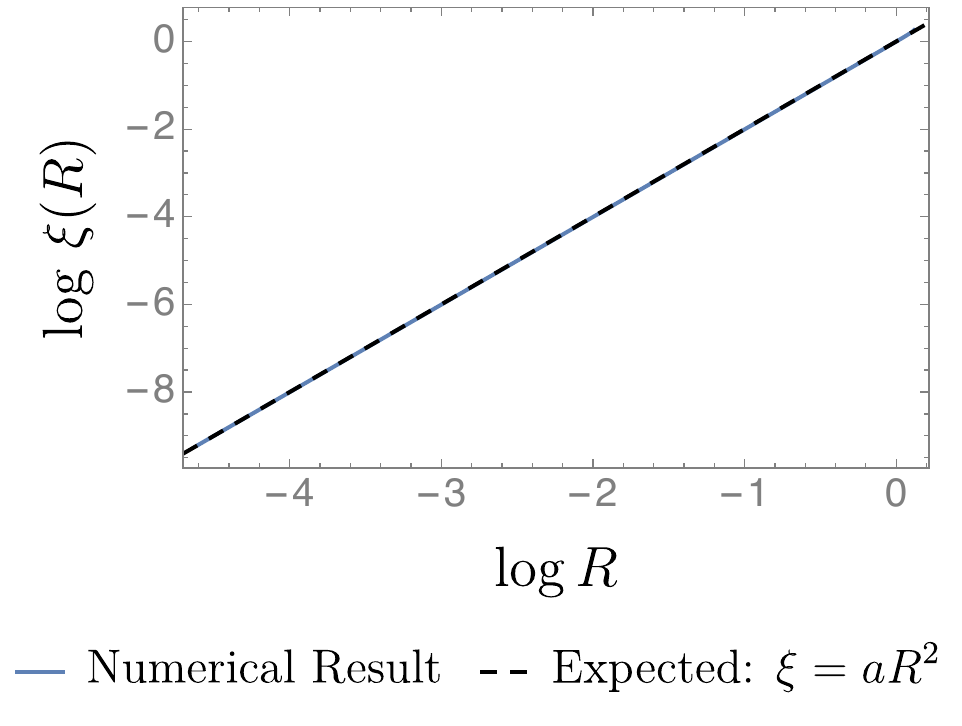}
  		\caption*{}		
	\end{subfigure}~
	\begin{subfigure}[t]{.33\textwidth}
		\centering
		\includegraphics[width=1.0\linewidth]{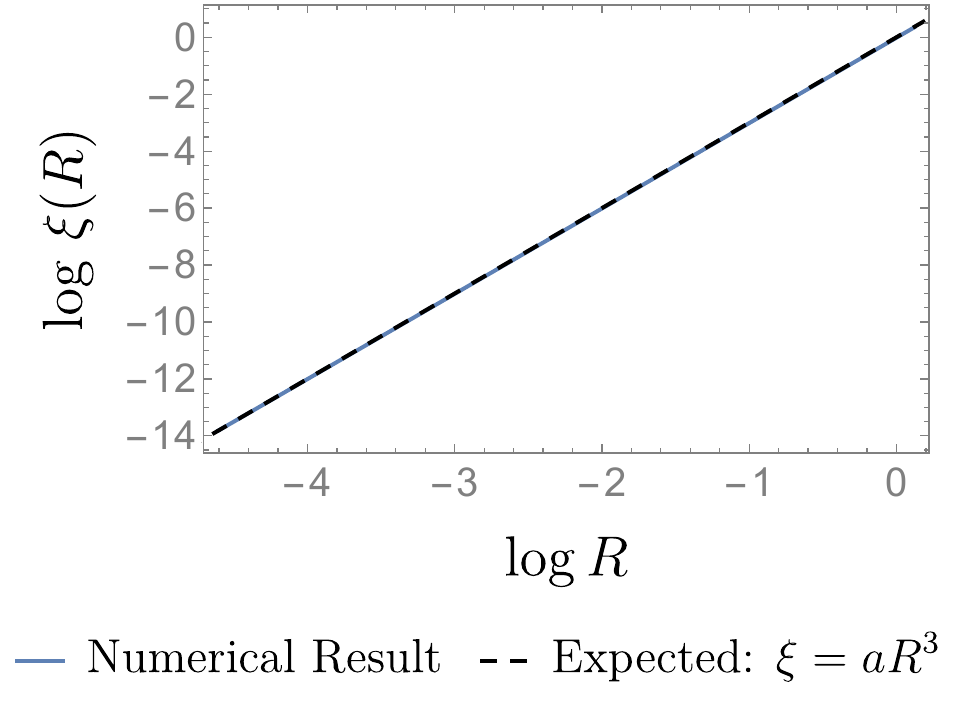}
  		\caption*{}		
	\end{subfigure}
	\caption{Comparison between the numerical result for the $k=\ell=m=0$ wavefunctions with $j=1,2,3$, corresponding to short states, and the expected analytical result: the modulus of (\ref{eq:EigenFunShort}) with (\ref{eq:Shortxi1}).}
	\label{fig: loglogplots}
\end{figure}

From the branching (\ref{eq:branchingSymTraceless}), the short graviton multiplets can be seen to correspond to bound states of the energy-momentum superfield and the operator ${\cal Z}^4$ that is integrated out in the IR. Schematically,
\begin{equation}	\label{eq: shortchain}
	\mathcal{T}^{\0}_{\alpha\beta} \big(\cZ^4 \big)^n  \; , \quad  n = 0 , 1 , 2 , \ldots\;,
\end{equation}
where $n=0$ corresponds to the massless graviton. Curiously, for the CPW geometry, the operators (\ref{eq: shortchain}) are also short \cite{Klebanov:2009kp}. Their physical properties remain as in (\ref{eq: massshorts})--(\ref{eq:ShortDelta}) with $R_2$ still given by the R-charge of ${\cal Z}^4 $, which now takes on the value (\ref{eq: Rassignmentp=2}). The group theory result (\ref{eq: shortchain}) is in agreement with our numerics, and in fact allows us to obtain the corresponding eigenfunctions analytically. The eigenfunction of (\ref{eq:spin2PDE}) with (\ref{eq:MassOpPart}), dual to the operator (\ref{eq: shortchain}), is given by 
\begin{equation} \label{eq:EigenFunShort}
	\cy_j=(\xi_1)^je^{ij\psi} \; ,
\end{equation}
in terms of $\xi_1(r)$, which is the $r$-dependent function $\xi_{\ell,m,j}(r)$ in (\ref{eq: modeexpansion}) with $j=1$, $\ell = m = 0$ and $k=0$ so that $n=1$ via (\ref{eq:QNredef}). The subscript in $\xi_1(r)$ refers to the fact that this function corresponds to an SU(3) singlet: the  SU(3) singlet at KK level $n=1$ in table \ref{tab:KKGravMassSU3U1N=2SO8}. Inserting the eigenfunction $\xi_1(r)$ and its analytic eigenvalue (\ref{eq: massshorts}) into (\ref{eq: ODExi}) with the above choice of quantum numbers, the ODE (\ref{eq: ODExi}) reduces to 
\begin{equation}	\label{eq: odeshort}
	(\xi'_1)^2=\frac1{9r^2}\xi_1^2\;.
\end{equation}
This equation can be analytically solved as
\begin{equation} \label{eq:Shortxi1}
\xi_1= r^{1/3} \equiv R \; , 
\end{equation}
in exact agreement with our numerical integration, see figure \ref{fig: loglogplots}. A similar analysis for CPW leads to $\xi_1= r^{1/2}$.

Our numerics strongly suggest that all other gravitons belong to long multiplets, with masses $M^2L^2$ leading to conformal dimensions $\Delta$ through (\ref{eq:DeltaM2}) that are above the bound (\ref{eq:ShortDelta}), $\Delta > |R | +3$. Group theory allows us to determine the structure of the dual operators as reported in table \ref{tab:KKGravMassSU3U1N=2SO8}, but in general we can only access the mass eigenvalues numerically. There is an exception: for a certain series of long gravitons starting at SO(8) KK level $n=2$, we can determine the masses analytically and relate the corresponding eigenfunctions to precise metric functions. These modes have $\textrm{SU}(3) \times \textrm{U}(1)_3$ charges $[1,1]_{\pm R_2   (n-2)} $, with $R_2$ again given in (\ref{eq: Rassignmentp=3}), and are dual to operators of the schematic form 
\begin{equation}	\label{eq: shadowchain}
	\mathcal{T}^{\0}_{\alpha\beta}\Big(\cZ^A\bZ_B-\tfrac13\delta^A_B\cZ^C\bZ_C\Big)(\cZ^4)^{n-2} \; , \quad n = 2 ,3 , \ldots 
\end{equation}
In \cite{Klebanov:2009kp} it was observed that the analogue tower of modes for CPW has dimensions
\begin{equation} \label{eq:DeltaShadow}
\Delta_n=(n-2)R_2+4
\end{equation}
(with $R_2$ accordingly given in (\ref{eq: Rassignmentp=2}) above). The authors of \cite{Klebanov:2009kp} suggested that this apparent protection of the conformal dimensions in terms of the R-charges for these modes occurs, despite being long, because they are {\it shadows} \cite{Billo:2000zs} of the massless vector at KK level $n=0$, which lies in the $\bm{8}_0$ of $\textrm{SU}(3) \times \textrm{U}(1)_2$.

\begin{figure}
\centering
	\begin{subfigure}[t]{.5\textwidth}
		\centering
		\includegraphics[width=1.0\linewidth]{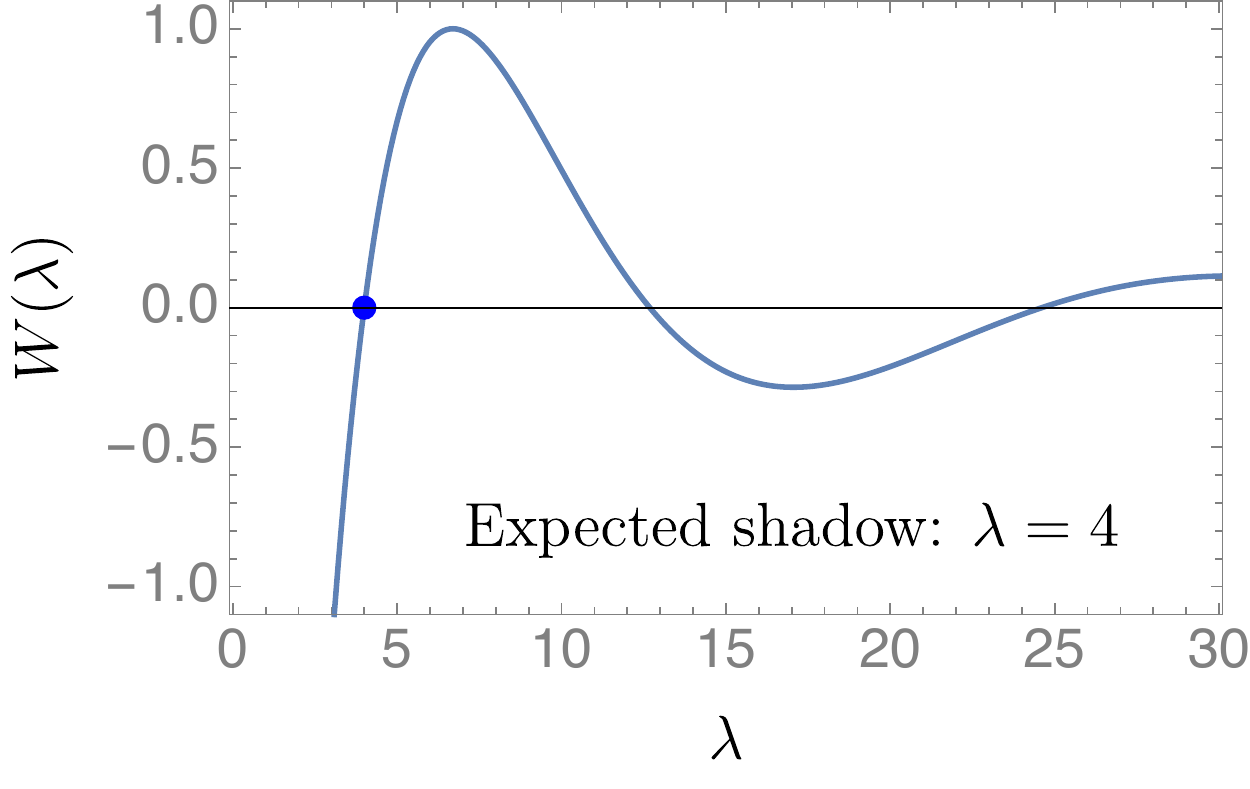}
  		\caption{}
  		\label{fig: wrl2}
	\end{subfigure}~
	\begin{subfigure}[t]{.5\textwidth}
		\centering
		\includegraphics[width=1.0\linewidth]{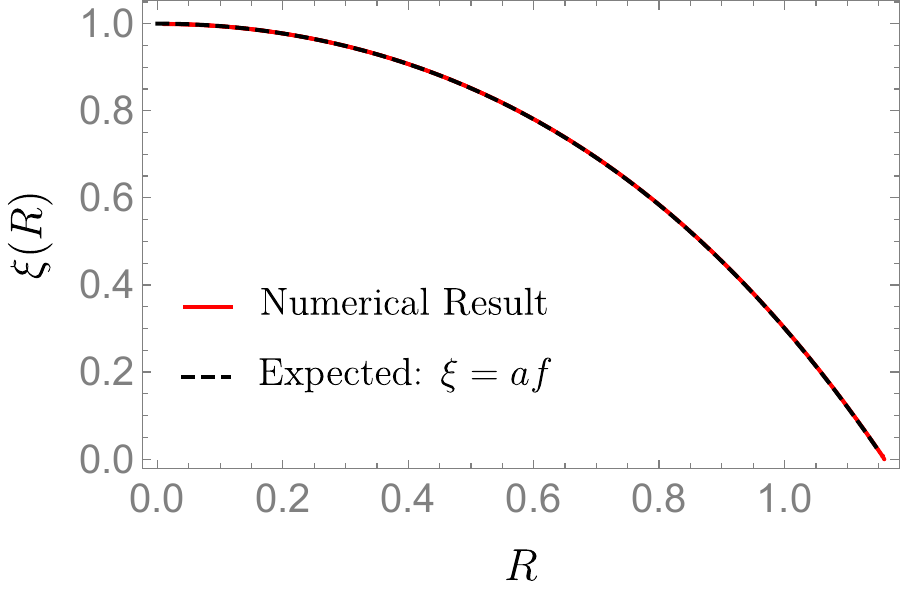}
  		\caption{}		
		\label{fig: wavefnl2k0}
	\end{subfigure}
	\caption{(a): Wronskian at $R=R_0/2$ of the functions $\xi^L_{\sst{\lambda}}(R)$ and $\xi^R_{\sst{\lambda}}(R)$ corresponding to shadow solutions with $\ell=2$, $m=j=0$. A blue dot signals the expected mass of a shadow octet state. (b): Wavefunction $\xi_8 (R)$ for the lightest shadow mode with $\ell=2$, $m=j=0$. The agreement of the numerical result  $\xi_8$ with the background function $a f$ is excellent, with the proportionality constant $a$ fixed to $a=2/9$.} \label{fig:Shadow}
\end{figure}

Our numerical routine described in section \ref{sec:NumBVP} finds a massive KK graviton over GMPS with quantum numbers $\ell=2$, $k=j=m=0$ and mass that can be very well approximated by the analytic value $L^2M^2=4$. In terms of the quantum numbers (\ref{ref:QNRanges}) associated to the branching (\ref{eq:branchingSymTraceless}), this state is attained at KK level $n=2$ with quantum numbers $\ell  = 2$, $p=1$, $t=0$. From (\ref{eq:DeltaM2}), the conformal dimension of this state is $\Delta = 4$, which agrees with (\ref{eq:DeltaShadow}) for $n=2$. This suggests that this state lies at the bottom of a tower of shadow gravitons with dual operators (\ref{eq: shadowchain}) and conformal dimensions (\ref{eq:DeltaShadow}), exactly as for CPW but now with $R_2$ given by (\ref{eq: Rassignmentp=3}). Our numerical integration confirms this expectation. We do find numerically a tower of masses that can be very well approximated by the analytic expression
\begin{equation} \label{eq:MassShadow}
L^2 M_n^2 =  \big( (n-2)R_2+4 \big) \big( (n-2)R_2+1 \big) \; , \qquad n = 2, 3 , \ldots 
\end{equation}
with $R_2$ as in (\ref{eq: Rassignmentp=3}). These masses indeed correspond to the conformal dimension (\ref{eq:DeltaShadow}) through (\ref{eq:DeltaM2}). 

For these shadow gravitons we can also relate their eigenfunctions to a precise metric function. The eigenfunctions (\ref{eq: modeexpansion}) corresponding to this tower of states can be written as
\begin{equation} \label{eq:EigenFunShadow}
	\cy_j= \xi_8\, r^{j/3}\, Y_{2,0} \, e^{ij\psi} \;,		\qquad j=0,1,\dots\;,
\end{equation}
where $\xi_8 (r) $ is the $r$-dependent part of the eigenfunction of the lightest state in the tower, with $\ell=2$, $k=j=m=0$. The subscript in $\xi_8(r)$ refers to the fact this function corresponds to an SU(3) octet: the  SU(3) octet, $[1,1]$, at KK level $n=2$ in table \ref{tab:KKGravMassSU3U1N=2SO8}. In (\ref{eq:EigenFunShadow}) we have assumed that the $(\cZ^4)^j$ contributions in (\ref{eq: shadowchain}) amount to factors of $(r^{1/3}e^{i\psi})^j$ in the eigenfunction by virtue of (\ref{eq:EigenFunShort}), (\ref{eq:Shortxi1}). The function $\xi_8$ satisfies the ODE (\ref{eq: ODExi}) for all $j$ and with the other quantum numbers suitably fixed, with mass eigenvalue (\ref{eq:MassShadow}) with $n$ there related to $j$ and $\ell=2$ through (\ref{eq:QNredef}). This discrete, $j$-dependent set of ODEs can be shown to be equivalent to the following set of two ODEs:
\begin{equation}	\label{eq: odeoctectsj0}
	\xi_8+\frac{2}{r\alpha^2}\xi_8'=0 \; , \qquad 
	\xi_8-\frac{3\sqrt{1+(1+r^2)\alpha^2}}{f\cdot\alpha}\xi_8+\frac{1}{r\alpha^2 f^3}\big(r f^3\xi'_8\big)'=0 \; .
\end{equation}
Now, the first ODE in (\ref{eq: odeoctectsj0}) is the same as the first of the ODEs in \eqref{eq: GstrRelations} that characterise the background geometry. We thus conclude that $\xi_8$ is proportional to the metric function $f$. Having used this proportionality, it can then be shown that the second ODE in (\ref{eq: odeoctectsj0}) can be deduced from \eqref{eq: GstrRelations}. The complete set of eigenfunctions for the tower of long shadow multiplets is thus given by (\ref{eq:EigenFunShadow}) with $\xi_8 \propto f$. See figure \ref{fig:Shadow}. We have verified that $\xi_8 \propto f$ also holds for the CPW case, with $f$ now given analytically in (\ref{eq:CPWMetricFuns}).

\subsection{GMPS is not isometrically embedded in $\mathbb{R}^8$} \label{sec:GMPSnotinR8}

\begin{figure}
	\centering
	\includegraphics[width=.6\linewidth]{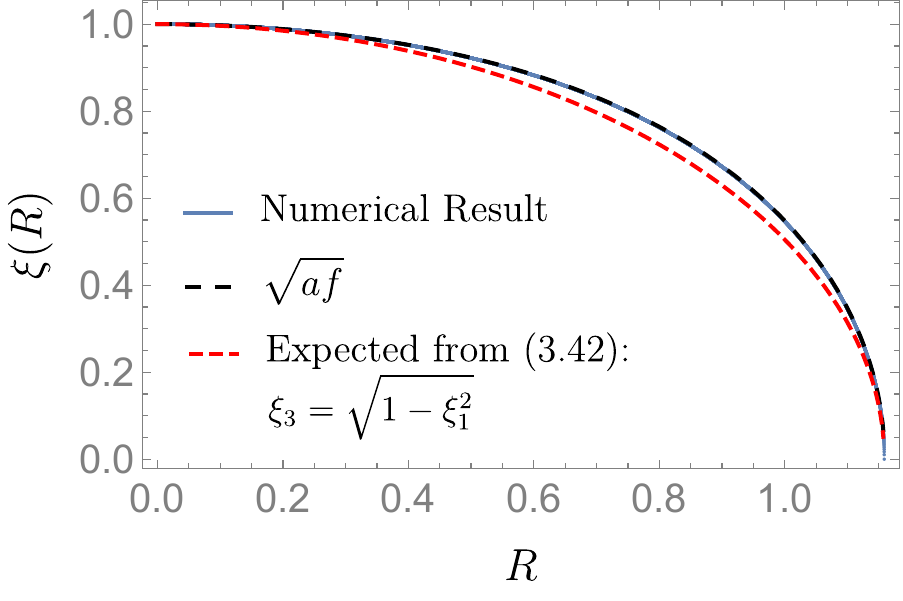}
  	\caption{ The radial wavefunction $\xi_3(R)$ of the triplet of long gravitons at KK level $n=1$. The numerically integrated result is matched by $\sqrt{f}$ up to a proportionality constant $a =2/9$, but not by the expression that would be expected if the $S^7$ constraint (\ref{eqS7Locus}) held.} 
	\label{fig:triplet}
\end{figure}

For all other long gravitons on the GMPS background, we do not have an argument to fix analytically their mass eigenvalues from our numerical results. Still, for the triplet, $[1,0]_{R_1}$,  of long gravitons at KK level $n=1$ we may ask whether the corresponding eigenfunction is $\xi_3 \propto \sqrt{f}$. This suspicion is based on the previous observation that $\xi_8 \propto f$, and that the radial part of the octet eigenfunction should be quadratically related to that of the triplet, in agreement with the group theory branching (\ref{eq:branchingSymTraceless}). Figure \ref{fig:triplet} shows that this is indeed the correct picture, as our numerically integrated $\xi_3$ perfectly matches $\sqrt{f}$ up to a numerical constant. Using the analytic expression (\ref{eq:CPWMetricFuns}), it is straightforward to check that $\xi_3 \propto \sqrt{f}$ also holds for the CPW solution.

It is also easy to verify for the CPW solution that the triplet, $\xi_3$, and singlet, $\xi_1$, radial eigenfunctions at KK level $n=1$ are related through the quadratic constraint that realises $S^7$ as a geometric locus in $\mathbb{R}^8$:
\begin{equation} \label{eqS7Locus}
\bar{Z}_C  Z^C + \bar{Z}_4 Z^4 = 1 \; .
 \end{equation}
Somewhat surprisingly, this relation does not hold for GMPS, as we will now show building on our results from section \ref{sec:ShortShadow}. To see this, let us assume (\ref{eqS7Locus}) and reach a contradiction. Equation (\ref{eqS7Locus}) implies 
\begin{equation} \label{eq:xi3Ansatz}
	\xi_3 \propto \sqrt{ 1- \xi_1^2} 
\end{equation}
by identifying the modulus of $Z^4$ with $\xi_1$ and that of $Z^C$ with $\xi_3$. Using $\xi_1= (r/r_0)^{1/3}$ as follows from a constant rescaling of (\ref{eq:Shortxi1}), $ \xi_3 \propto \sqrt{f}$ as verified in figure \ref{fig:triplet}, and $\xi_ 8 \propto \xi_3^2 \propto f$ as shown in figure \ref{fig:Shadow}, we conclude from (\ref{eq:xi3Ansatz}) that
\begin{equation} \label{eq:xi8Ansatz}
	\xi_8\propto1-\Big(\frac r{r_0}\Big)^{2/3}	
\end{equation}
for the octet at level $n=2$. Using (\ref{eq: odeoctectsj0}), we finally manage to obtain the following explicit expression for the $\alpha$ metric function: 
\begin{equation} \label{eq:PosAlphaGMPS}
	\alpha^2=\frac{4}{3 r^2 \Big[\big(\frac{r}{r_0}\big)^{-2/3}-1\Big]} \; .
\end{equation}
Remarkably, this expression obeys the correct asymptotics \eqref{eq: asymptotics}. 
Unfortunately, the function $\alpha$ in (\ref{eq:PosAlphaGMPS}) does not satisfy the second ODE in \eqref{eq: GstrRelations} for any value of $r_0$ and thus cannot be the correct GMPS metric function. In contrast, for CPW the same logic starting from (\ref{eqS7Locus}) allows one to recover the correct $\alpha$ in (\ref{eq:CPWMetricFuns}). The failure of the argument for GMPS leads us to abandon the hypothesis that (\ref{eqS7Locus}) should hold in the latter case. Equations (\ref{eq:xi3Ansatz})--(\ref{eq:PosAlphaGMPS}) for GMPS are false, as must be the original assumption (\ref{eqS7Locus}). Indeed, figure \ref{fig:triplet} manifestly shows that (\ref{eq:xi3Ansatz}) as  derived from the $S^7$ contraint (\ref{eqS7Locus}) does not reproduce our numerical result for $\xi_3$, not even including a proportionality constant.

From this discussion, we infer that the GMPS geometry is defined on a topological $S^7$ that, however, fails to satisfy the relation (\ref{eqS7Locus}) and thus is not embedded isometrically in $\mathbb{R}^8$. Another example of an $\textrm{AdS}_4 \times S^7$ solution for which (\ref{eqS7Locus}) does not hold is provided by the squashed $S^7$ of \cite{Awada:1982pk}. This has consequences for the general KK spectrum, as we now turn to discuss.


\section{Space invaders  scenario } \label{sec: spaceinvaders}


We would like to conclude with some comments about the full KK spectrum over the GMPS solution \cite{ Gabella:2012rc}. It is certainly beyond the scope of this work to compute the full spectrum. Instead, we will content ourselves with drawing  some conclusions  from group theory about its structure, as similarly done in \cite{Klebanov:2008vq} for the CPW solution \cite{Corrado:2001nv}. The main observation is that the KK spectrum displays a {\it space invaders scenario} similar to that described in \cite{Duff:1986hr} for the KK spectrum on the 
squashed $S^7$ solution \cite{Awada:1982pk}.

As remarked in section \ref{sec:ShortShadow}, the full (super)symmetry group of the GMPS solution is OSp$(4|2) \times \textrm{SU}(3)$, and the KK spectrum must accordingly organise itself in representations of this (super)group. See appendix A of \cite{Klebanov:2008vq} for a convenient summary of OSp$(4|2)$ multiplets. In section \ref{sec:ShortShadow}, we branched the SO(8) KK graviton representation $G_n \equiv [n,0,0,0]$ at KK level $n$ under the internal bosonic symmetry group decomposition (\ref{eq:SO8toSU3U1}) to find the $\textrm{SU}(3) \times \textrm{U}(1)_3$ graviton charges at each KK level. In appendix \ref{sec: branchings}, we have performed this exercise starting from all other SO(8) towers for every KK level $n$:
\begin{eqnarray} \label{eq:SO8Irreps}
\textrm{graviton} &: & G_n  \equiv  [n,0,0,0] \; , \nonumber \\[4pt]
\textrm{gravitini} &: & \mathcal{G}_n  \equiv  [n,0,0,1]\oplus[n-1,0,1,0]  \; , \nonumber \\[4pt]
\textrm{vectors} &: & V_n  \equiv  [n,1,0,0]\oplus[n-1,0,1,1]\oplus[n-2,1,0,0]  \; , \nonumber \\[4pt] 
\textrm{fermions} &: & \mathcal{F}_n \equiv [n+1,0,1,0]\oplus[n-1,1,1,0]\oplus[n-2,1,0,1]\oplus[n-2,0,0,1 ]  \; , \nonumber \\[4pt] 
\textrm{scalars} &: & S^+_n  \equiv  [n+2,0,0,0]\oplus[n-2,2,0,0]\oplus[n-2,0,0,0]  \; , \nonumber \\[4pt] 
\textrm{pseudoscalars} &: & S^-_n  \equiv  	[n,0,2,0]\oplus[n-2,0,0,2]  \; , 
\end{eqnarray}
where only representations with non-negative Dynkin labels contribute at level $n$. Branching the SO(8) representations (\ref{eq:SO8Irreps}) under (\ref{eq:SO8toSU3U1}) as in appendix \ref{sec: branchings}, we determine how the full KK spectrum arranges itself in terms of $\textrm{SU}(3) \times \textrm{U}(1)_3$ representations.


\begin{table}[t]
\begin{center}
\resizebox{1.05\textwidth}{!}{
\begin{tabular}{|c|c|c|c|c|c|c|c|c|c|c|c|c|c|} \hline
Spin & SO(8) & \multicolumn{12}{c|}{$\text{SU}(3)\times\text{U}(1)_3$} \\ \hline
%
%
$2$ 	&	$\bf{1}$ & 		$\bm{1}_{0}$ &&&&&&&&&&&   \\ \hline
%
%
$\frac{3}{2}$	&	$\bm{8}_s$	& $\bm{1}_{+1}$ 	& $\bm{3}_{\frac19}$ & $\bm{\bar{3}}_{-\frac19}$ &  &  &&& &&&&\\
&&$\bm{1}_{-1}$ 	&&&&&&&&&&&  \\  \hline
%
%
$1$ & $\bm{28}$ &  $\bm{1}_{0}$ & $\bm{3}_{-\frac89}$ & $\bm{\bar{3}}_{\frac89}$& $\bm{8}_{0}$ &$\bm{3}_{-\frac29}$&$\bm{\bar{3}}_{\frac29}$&$\bm{1}_0$&&&&& \\
&&& $\bm{3}_{\frac{10}9}$ & $\bm{\bar{3}}_{-\frac{10}9}$ &&&&&&&&& \\\cline{2-14}
&\Rocket&& $\bm{3}_{-\frac89}$ & $\bm{\bar{3}}_{\frac89}$  && &&&&&&& \\ \hline
%
%
$\frac{1}{2}$ & $\bm{56}_s$ &   & $\bm{3}_{\frac19}$ & $\bm{\bar{3}}_{-\frac19}$ & $\bm{8}_{+1}$ &$\bm{3}_{\frac79}$& $\bm{\bar{3}}_{-\frac79}$ & $\bm{1}_{+1}$ &$\bm{6}_{-\frac19}$
	&$\bm{\bar{6}}_{\frac19}$&$\bm{1}_{\frac13}$&$\bm{1}_{-\frac13}$&  \\
&&&  $\bm{3}_{\frac19}$ & $\bm{\bar{3}}_{-\frac19}$ & $\bm{8}_{-1}$ & $\bm{3}_{-\frac{11}9}$ & $\bm{\bar{3}}_{\frac{11}9}$  & $\bm{1}_{-1}$ &&&&& \\\cline{2-14}
&\Rocket& & $\bm{3}_{-\frac{17}9}$  & $\bm{\bar{3}}_{\frac{17}9}$ && $\bm{3}_{\frac79}$ & $\bm{\bar{3}}_{-\frac79}$ & $\bm{1}_{+1}$ &&&&&$\bm{3}_{\frac19}$	\\
&&&&&&&& $\bm{1}_{-1}$ &&&&&$\bm{\bar{3}}_{-\frac19}$ \\ \hline

%
%
$0$ & $\bm{35}_v$ &  $$ & $$ & $$& $\bm{8}_{0}$ & $\bm{3}_{-\frac29}$ & $\bm{\bar{3}}_{\frac29}$ & $\bm{1}_{0}$ &$\bm{6}_{\frac89}$&$\bm{\bar{6}}_{-\frac89}$&$\bm{1}_{\frac43}$
	&$\bm{1}_{-\frac43}$& $\bm{3}_{\frac{10}9}$ \\
&&& $$ & $$ &&&&&&&&& $\bm{\bar{3}}_{-\frac{10}9}$\\\cline{2-14}
%
%
& $\bm{35}_c$ &  $$ & $\bm{3}_{-\frac89}$ & $\bm{\bar{3}}_{\frac89}$& $\bm{8}_{0}$ & $\bm{3}_{-\frac29}$ & $\bm{\bar{3}}_{\frac29}$ & $\bm{1}_{0}$ &$\bm{6}_{-\frac{10}9}$
	&$\bm{\bar{6}}_{\frac{10}9}$&$\bm{1}_{-\frac23}$&$\bm{1}_{\frac23}$& \\
&&& $$ & $$ &&&&&&&&& \\\cline{2-14}
&\Rocket&& $$ & $$  && $\bm{3}_{\frac{16}9}$ & $\bm{\bar{3}}_{-\frac{16}9}$ & $\bm{1}_{0}$ &&&&& $\bm{3}_{-\frac89}$, $\bm{\bar{3}}_{\frac89}$ \\ 
&&&&&&&& $\bm{1}_{+2}$ &&&&&  $\bm{3}_{-\frac89}$, $\bm{\bar{3}}_{\frac89}$ \\
&&&&&&&& $\bm{1}_{-2}$ &&&&&  $\bm{3}_{-\frac29}$, $\bm{\bar{3}}_{\frac29}$\\ 
&&&&&&&&&&&&& $\bm{1}_0$ \\ \hline

\multicolumn{2}{l|}{} &
\rotatebox{90}{\mbox{Massless graviton\;}} &
\rotatebox{90}{\mbox{Short gravitino\;}} &
\rotatebox{90}{\mbox{Short gravitino\;}} &
\rotatebox{90}{\mbox{Massless vector\;}} &
\rotatebox{90}{\mbox{Short vector\;}} &
\rotatebox{90}{\mbox{Short vector\;}} &
\rotatebox{90}{\mbox{Long vector\;}} &
\rotatebox{90}{\mbox{Massive hyper\;}} &
\rotatebox{90}{\mbox{Massive hyper\;}} &
\rotatebox{90}{\mbox{Massive hyper\;}} &
\rotatebox{90}{\mbox{Massive hyper\;}} &
\rotatebox{90}{\mbox{Eaten modes\;}} \\  \cline{3-14}
\end{tabular}
}
\caption{Possible branching of the $\mathcal{N}=8$ massless graviton multiplet into Osp$(4\vert2) \times \textrm{SU}(3)$ representations. The symbol \protect\Rocket\ denotes space invader states coming from KK level $n=1$. See table \ref{tab:multipletsatlevel0} in appendix \ref{sec:nAndn+1Pattern} for a summary.}
\label{table: branchKK0}
\end{center}
\end{table}

The next step is to allocate fields of different spin but the same $\textrm{SU}(3)$ charges into OSp$(4|2)$ multiplets. For CPW \cite{Corrado:2001nv} this exercise was carried out in \cite{Klebanov:2008vq}, and crucially relies on the assignment of R-charges (\ref{eq: Rassignmentp=2}). Under the assumption that the allocation into supermultiplets should take place KK level by KK level, group theory alone was found to narrow down the possible spectrum of (short) multiplets to two possibilities dubbed scenarios I and II in \cite{Klebanov:2008vq}. Both scenarios differ by the embedding of the U$(1)_2$ IR isometry into SO(8), and are related by a triality rotation \cite{Pang:2017omp}. The actual calculation of the KK graviton spectrum \cite{Klebanov:2009kp} confirmed scenario I as the correct choice.

Going through the same exercise for the GMPS solution \cite{Gabella:2012rc} we find that we need to relax the  assumption that the allocation of $\textrm{SU}(3) \times \textrm{U}(1)_3$ states into OSp$(4|2)$ multiplets should proceed KK level by KK level. Otherwise, the problem has no solution starting from the R-charge assignment (\ref{eq: Rassignmentp=3}), and that is not an option. Instead, states entering the same OSp$(4|2)$ multiplet must be retrieved from different SO(8) KK levels $n$. For example, states from higher KK levels are needed to complete Short Gravitino multiplets in the $[1,0]_{\frac19}$ and $[1,0]_{-\frac19}$ and a Long Vector in the $[0,0]_{0}$, whose states come mostly from $n=0$. Table~\ref{table: branchKK0} shows a possible distribution of the $n=0$ states into OSp$(4|2)$ multiplets that assumes that all needed space invaders descend from KK level $n=1$. Group theory is not enough to determine whether this or another invasion pattern is the correct one, though. Appendix \ref{app: scenarios} further speculates about this and other invasion patterns.

The other example we are aware of where a similar space invaders scenario occurs \cite{Duff:1986hr} is the $D=11$ AdS$_4$ solution  based on the squashed $S^7$ of \cite{Awada:1982pk}. Some features are common to GMPS and the squashed $S^7$ that lead to the existence of a space invaders scenario for their KK spectra. Firstly, neither of them arises as a vacuum of a consistently truncated $D=4$ $\cN=8$ supergravity. Secondly, while both solutions are defined on a topological $S^7$, their metrics cannot be isometrically embedded in $\mathbb{R}^8$ via (\ref{eqS7Locus}). The squashed $S^7$ metric is instead embedded in the quaternionic projective space $\mathbb{HP}^2$ \cite{Duff:1986hr}. It would be interesting to determine if, similarly, GMPS could be embedded into the complex projective space $\mathbb{CP}^4$ by appropriately embedding the isometry group SU(3). A notable difference between the squashed $S^7$ and GMPS is that the former is homogeneous while the latter is cohomogeneity-one. This feature allowed the authors of \cite{Nilsson:2018lof} to compute the complete KK spectrum on the squashed $S^7$ using techniques relevant to homogeneous spaces, which are obviously unavailable for GMPS. The spectrum generating technique of \cite{Malek:2019eaz} (see also \cite{Dimmitt:2019qla}) is not readily available either, as it relies on the existence of an $\cN=8$ consistent truncation. It would be interesting to investigate if some modification of these techniques allows for the computation of the complete KK spectrum over the GMPS solution.


\section*{Acknowledgements}


We have found inspiration in the title of \cite{Klebanov:2009kp} for the subtitle of this paper. We would like to thank Daniel Jafferis for conversations and Praxitelis Ntokos for collaboration in the early stages of this work. MC is supported by a La Caixa Foundation (ID 100010434) predoctoral fellowship LCF/ BQ/DI19/11730027. GL is supported by an FPI-UAM predoctoral fellowship. OV is supported by the NSF grant PHY-1720364. MC, GL and OV are partially sup\-por\-ted by grants SEV-2016-0597 and PGC2018-095976-B-C21 from MCIU/AEI/FEDER, UE.

\appendix

\addtocontents{toc}{\setcounter{tocdepth}{1}}



\section{Embedding SU(3)$\times$U(1)$_p$ into SO(8)} \label{sec: branchings}
\addtocontents{toc}{\setcounter{tocdepth}{1}}


The internal bosonic symmetry group SU(3)$\times$U(1)$_p$, with $p = 2 $ for CPW and $p=3$ for GMPS, is embedded into SO(8) via
\begin{equation} \label{eq:SO8intoSU3}
	\text{SO(8)}\supset\text{SO(6)}_v \times\text{SO(2)}\supset \big[ \text{SU(3)}\times\text{U(1)} \big] \times\text{SO(2)}\supset\text{SU(3)}\times\text{U(1)}_{p}\; .
\end{equation}
Under the first two steps in the branching (\ref{eq:SO8intoSU3}), the three basic irreps of SO(8) split as
\begin{eqnarray}	\label{eq: 8sofSO8toU3branching}
	&& \bm{8}_v \;
	\longrightarrow	\;
	\ \makebox[2cm][c]{ $\bm{6}_0+\bm{1}_1+\bm{1}_{-1}$ }\ 
	\longrightarrow \;
	\ \makebox[6cm][c]{ $\bm{3}_{(-\frac23,\;0)} +  \overline{\bm{3}}_{(\frac23,\;0)} + \bm{1}_{(0,\;+1)}  +  \bm{1}_{(0,\;-1)}$ }  \; ,
	\nonumber\\[7pt]
	&& \bm{8}_s \;
	\longrightarrow	\;
	\ \makebox[2cm][c]{ $\bm{4}_{\frac12}+\overline{\bm{4}}_{-\frac12}$ }\ 
	\longrightarrow \;
	\ \makebox[6.5cm][c]{ $\bm{3}_{(\frac13,\;\frac12)} +  \overline{\bm{3}}_{(-\frac13,\;-\frac12)} + \bm{1}_{(-1,\;\frac12)}  +  \bm{1}_{(+1,\;-\frac12)}$ }  \; , 
	\nonumber\\[7pt]
	&& \bm{8}_c \;
	\longrightarrow	\;
	\ \makebox[2cm][c]{ $\bm{4}_{-\frac12}+\overline{\bm{4}}_{\frac12}$ }\ 
	\longrightarrow \;
	\ \makebox[6.5cm][c]{ $\bm{3}_{(\frac13,\;-\frac12)} +  \overline{\bm{3}}_{(-\frac13,\;\frac12)} + \bm{1}_{(-1,\;-\frac12)}  +  \bm{1}_{(+1,\;\frac12)}$ }  \; .
\end{eqnarray}
The IR R-symmetry group U(1)$_p$ is the combination of the U$(1)$ that commutes with SU(3) inside SO$(6)_v$ and the SO(2) that commutes with SO$(6)_v$ inside SO(8) which leads to the allocation of R-charges \eqref{eq: Rassignmentp=3} for $p=3$ and \eqref{eq: Rassignmentp=2} for $p=2$. Assigning the transverse M2-brane coordinates to the $\mathbf{8}_v$, we thus require that, under the third and final step in the branching (\ref{eq:SO8intoSU3}),
\begin{equation} \label{eq: 8sofSO8toSU3U1v}
	\bm{8}_v	\longrightarrow \bm{3}_{R_1}+\overline{\bm{3}}_{-R_1}+\bm{1}_{R_2}+\bm{1}_{-R_2}\; ,
\end{equation}
with 
\begin{equation} \label{eq:Rcharges}
R_1  = \frac{2p-2}{3p} \; , \qquad
R_2  = \frac{2}{p} \; .
\end{equation}
For completeness, we note that
\begin{eqnarray}	\label{eq: 8sofSO8toSU3U1sc}
	&& \bm{8}_s \;
	\longrightarrow	\;
	\bm{3}_{\tfrac12(-R_1+R_2)}+\overline{\bm{3}}_{\tfrac12(R_1-R_2)}+\bm{1}_{\tfrac12(3R_1+R_2)}+\bm{1}_{-\tfrac12(3R_1+R_2)}	\;,
	\nonumber\\[7pt]
	&& \bm{8}_c \;
	\longrightarrow	\;
	\bm{3}_{-\tfrac12(R_1+R_2)}+\overline{\bm{3}}_{\tfrac12(R_1+R_2)}+\bm{1}_{\tfrac12(3R_1-R_2)}+\bm{1}_{\tfrac12(-3R_1+R_2)}	\;.
\end{eqnarray}

Taking tensor products and (anti)symmetrisations of (\ref{eq: 8sofSO8toSU3U1v}), (\ref{eq: 8sofSO8toSU3U1sc}), an arduous calculation allows us to find the branching under SU(3)$\times$U(1)$_p$ of the SO(8) representations (\ref{eq:SO8Irreps}) that characterise the KK spectrum at the $\cN=8$ point. We obtain\footnote{In (A.6)--(A.11) we have renamed $\textrm{SU}(3)\times \textrm{U}(1)_p$ as $\textrm{SU}(3)\times \textrm{U}(1)_\textrm{R}$ in order to avoid confusion with the Dynkin label $p$. }
\begin{eqnarray}
	G_n	\!\!\!\!\!\!\! &&=\ 	[n,0,0,0]	\nonumber\\
				&&\xrightarrow{{\scriptsize \text{SU}(3)\times\text{U}(1)_{\text{R}}}} \bigoplus_{\ell=0}^{n}\bigoplus_{t=0}^{n-\ell}\bigoplus_{p=0}^{\ell}
				[p,\;\ell-p]_{-R_1(\ell-2p)+R_2(n-\ell-2t)}\;,
\end{eqnarray}

\newpage 

\begin{eqnarray}
	\mathcal{G}_n	\!\!\!\!\!\!\! &&=\ 	[n,0,0,1]\oplus[n-1,0,1,0]	\nonumber\\[7pt]
				&&\xrightarrow{{\scriptsize \text{SU}(3)\times\text{U}(1)_{\text{R}}}} \nonumber\\[6pt]
				&&\bigoplus_{\ell=0}^{n}\bigoplus_{t=0}^{n-\ell}\bigoplus_{p=0}^{\ell}\bigoplus_{k=0}^{1}\bigoplus_{a=0}^{1-k}\bigoplus_{b=0}^{k}\;
				[p+1-k-a,\;\ell-p+k-b]_{\substack{-R_1(\ell-2p-k-2a+2b+\frac12)\\+R_2(n-\ell-2t-k+\frac12)}}\nonumber\\
				&&\oplus\bigoplus_{\ell=0}^{n-1}\bigoplus_{t=0}^{n-1-\ell}\bigoplus_{p=0}^{\ell}\bigoplus_{k=0}^{1}\bigoplus_{a=0}^{1-k}\bigoplus_{b=0}^{k}\;
				[p+1-k-a,\;\ell-p+k-b]_{\substack{-R_1(\ell-2p-k-2a+2b+\frac12)\\+R_2(n-\ell-2t+k-\frac32)}}\;,\nonumber\\
\end{eqnarray}
\begin{eqnarray}
	V_n	\!\!\!\!\!\!\! &&=\ 	[n,1,0,0]\oplus[n-1,0,1,1]\oplus[n-2,1,0,0]	\nonumber\\[7pt]
				&&\xrightarrow{{\scriptsize \text{SU}(3)\times\text{U}(1)_{\text{R}}}} \nonumber\\[6pt]
				&&\bigoplus_{\ell=0}^{n}\bigoplus_{t=0}^{n-\ell}\bigoplus_{p=0}^{\ell}\bigoplus_{a,b=0}^{1}\;
				[p+a,\;\ell-p+b]_{\substack{-R_1(\ell-2p+2a-2b)\\+R_2(n-\ell-2t)}}\nonumber\\
				&&\quad\oplus\bigoplus_{\ell=0}^{n}\bigoplus_{t=0}^{n-\ell}\bigoplus_{p=0}^{\ell+1}\bigoplus_{k=0}^{1}\;
				[p,\;\ell-p+1]_{\substack{-R_1(\ell-2p+1)\\+R_2(n-\ell-2t-2k+1)}}
				\oplus\bigoplus_{\ell=0}^{n}\;[0,0]_{R_2(n-2\ell)}\nonumber\\
				&&\oplus\bigoplus_{\ell=0}^{n-1}\bigoplus_{t=0}^{n-1-\ell}
				\bigoplus_{p=0}^{\ell}\bigoplus_{a,b=0}^1\bigoplus_{c=0}^{2-a-b}\bigoplus_{d=0}^{a+b}\;[p+c,\,\ell-p+d]_{\substack{-R_1(\ell-2p+3a+3b+2c-2d-3)\\+R_2(n-\ell-2t+a-b-1)}}\nonumber\\
				&&\quad\oplus\bigoplus_{\ell=0}^{n-1}\bigoplus_{t=0}^{n-1-\ell}\bigoplus_{p=0}^{\ell+1}\;[p,\;\ell-p+1]_{\substack{-R_1(\ell-2p+1)\\+R_2(n-\ell-2t-1)}}\nonumber\\
				&&\oplus\bigoplus_{\ell=0}^{n-2}\bigoplus_{t=0}^{n-2-\ell}\bigoplus_{p=0}^{\ell}\bigoplus_{a,b=0}^{1}\;
				[p+a,\;\ell-p+b]_{\substack{-R_1(\ell-2p+2a-2b)\\+R_2(n-\ell-2t-2)}}\nonumber\\
				&&\quad\oplus\bigoplus_{\ell=0}^{n-2}\bigoplus_{t=0}^{n-2-\ell}\bigoplus_{p=0}^{\ell+1}\bigoplus_{k=0}^{1}\;
				[p,\;\ell-p+1]_{\substack{-R_1(\ell-2p+1)\\+R_2(n-\ell-2t-2k-1)}}
				\oplus\bigoplus_{\ell=0}^{n-2}\;[0,0]_{R_2(n-2\ell-2)}\;,\nonumber\\
\end{eqnarray}

\newpage 

\begin{eqnarray}
	\mathcal{F}_n	\!\!\!\!\!\!\! &&=\ 	[n+1,0,1,0]\oplus[n-1,1,1,0]\oplus[n-2,1,0,1]\oplus[n-2,0,0,1]	\nonumber\\[7pt]
				&&\xrightarrow{{\scriptsize \text{SU}(3)\times\text{U}(1)_{\text{R}}}} \nonumber\\[6pt]
				&&\bigoplus_{\ell=0}^{n+1}\bigoplus_{t=0}^{n+1-\ell}\bigoplus_{p=0}^{\ell}\bigoplus_{k=0}^{1}\bigoplus_{a=0}^{1-k}\bigoplus_{b=0}^{k}\;
				[p+1-k-a,\;\ell-p+k-b]_{\substack{-R_1(\ell-2p-k-2a+2b+\frac12)\\+R_2(n-\ell-2t+k+\frac12)}}		\nonumber\\
				&&\oplus\bigoplus_{\ell=0}^{n-1}\bigoplus_{t=0}^{n-1-\ell}\bigoplus_{q=0}^{1}\bigoplus_{p=0}^{\ell}\bigoplus_{a=0}^{q+1}\bigoplus_{b=0}^{2-q}\;
				[p+a,\; \ell-p+b]_{\substack{-R_1(\ell-2p-3q+2a-2b+\frac32)\\+R_2(n-\ell-2t-q-\frac12)}}	\nonumber\\
				&&\quad\oplus\bigoplus_{\ell=0}^{n-1}\bigoplus_{t=0}^{n-1-\ell}\bigoplus_{k,q=0}^{1}\bigoplus_{p=0}^{\ell+1}\bigoplus_{a=0}^{q}\bigoplus_{b=0}^{1-q}\;
				[p+a,\; \ell+1-p+b]_{\substack{-R_1(\ell-2p-3q+2a-2b+\frac52)\\+R_2(n-\ell-2t-2k-q+\frac12)}}	\nonumber\\
				&&\qquad\oplus\bigoplus_{\ell=0}^{n-1}\bigoplus_{q=0}^{1}\bigoplus_{a=0}^{q}\bigoplus_{b=0}^{1-q}\;
				[a,\; b]_{-R_1(\frac32-3q+2a-2b)+R_2(n-2\ell-q-\frac12)}	\nonumber\\
				&&\oplus\bigoplus_{\ell=0}^{n-2}\bigoplus_{t=0}^{n-2-\ell}\bigoplus_{q=0}^{1}\bigoplus_{p=0}^{\ell}\bigoplus_{a=0}^{q+1}\bigoplus_{b=0}^{2-q}\;
				[p+a,\; \ell-p+b]_{\substack{-R_1(\ell-2p-3q+2a-2b+\frac32)\\+R_2(n-\ell-2t+q-\frac52)}}	\nonumber\\
				&&\quad\oplus\bigoplus_{\ell=0}^{n-2}\bigoplus_{t=0}^{n-2-\ell}\bigoplus_{k,q=0}^{1}\bigoplus_{p=0}^{\ell+1}\bigoplus_{a=0}^{q}\bigoplus_{b=0}^{1-q}\;
				[p+a,\; \ell+1-p+b]_{\substack{-R_1(\ell-2p-3q+2a-2b+\frac52)\\+R_2(n-\ell-2t+2k+q-\frac72)}}	\nonumber\\
				&&\qquad\oplus\bigoplus_{\ell=0}^{n-2}\bigoplus_{q=0}^{1}\bigoplus_{a=0}^{q}\bigoplus_{b=0}^{1-q}\;
				[a,\; b]_{-R_1(\frac32-3q+2a-2b)+R_2(n-2\ell+q-\frac52)}	\nonumber\\
				&&\oplus\bigoplus_{\ell=0}^{n-2}\bigoplus_{t=0}^{n-2-\ell}\bigoplus_{p=0}^{\ell}\bigoplus_{k=0}^{1}\bigoplus_{a=0}^{1-k}\bigoplus_{b=0}^{k}\;
				[p+1-k-a,\;\ell-p+k-b]_{\substack{-R_1(\ell-2p-k-2a+2b+\frac12)\\+R_2(n-\ell-2t-k-\frac32)}}\;,\nonumber\\
\end{eqnarray}
\begin{eqnarray}
	S^+_n\!\!\!\!\!\!\! &&=\ 	[n+2,0,0,0]\oplus[n-2,2,0,0]\oplus[n-2,0,0,0]	\nonumber\\[7pt]
				&&\xrightarrow{{\scriptsize \text{SU}(3)\times\text{U}(1)_{\text{R}}}} \nonumber\\[6pt]
				&&\bigoplus_{\ell=0}^{n+2}\bigoplus_{t=0}^{n+2-\ell}\bigoplus_{p=0}^{\ell}\;
				[p,\;\ell-p]_{-R_1(\ell-2p)+R_2(n-\ell-2t+2)}	\nonumber\\
				&&\oplus\bigoplus_{\ell}^{n-2}\bigoplus_{t=0}^{n-2-\ell}\bigoplus_{q=0}^2\bigoplus_{k=0}^q\bigoplus_{p=0}^{\ell+q}\bigoplus_{a,b=0}^{2-q}\;
				[p+a,\;\ell+q-p+b]_{\substack{-R_1(\ell+q-2p+2a-2b)\\+R_2(n-\ell-2t+q-2k-2)}}	\nonumber\\
				&&\quad\oplus\bigoplus_{\ell}^{n-2}\bigoplus_{k=0}^1\bigoplus_{a,b=0}^k\;
				[a,\;b]_{\substack{-2R_1(a-b)\\+R_2(n-2\ell-2)}}
				\oplus\bigoplus_{\ell}^{n-2}\bigoplus_{k,p=0}^1\;
				[p,\;1-p]_{\substack{-R_1(1-2p)\\+R_2(n-2\ell-2k-1)}}	\nonumber\\
				&&\oplus\bigoplus_{\ell=0}^{n-2}\bigoplus_{t=0}^{n-2-\ell}\bigoplus_{p=0}^{\ell}\;
				[p,\;\ell-p]_{-R_1(\ell-2p)+R_2(n-\ell-2t-2)}\;,
\end{eqnarray}
\begin{eqnarray}
	S^-_n\!\!\!\!\!\!\! &&=\ 	[n,0,2,0]\oplus[n-2,0,0,2]	\nonumber\\[7pt]
				&&\xrightarrow{{\scriptsize \text{SU}(3)\times\text{U}(1)_{\text{R}}}} \nonumber\\[6pt]
				&&\bigoplus_{\ell=0}^{n}\bigoplus_{t=0}^{n-\ell}\bigoplus_{k=0}^{2}\bigoplus_{p=0}^{\ell}\bigoplus_{a=0}^{2-k}\bigoplus_{b=0}^{k}\;
				[p+a,\;\ell-p+b]_{\substack{-R_1(\ell-2p+3k+2a-2b-3)\\+R_2(n-\ell-2t+k-1)}}	\nonumber\\
				&&\oplus\bigoplus_{\ell=0}^{n-2}\bigoplus_{t=0}^{n-2-\ell}\bigoplus_{k=0}^{2}\bigoplus_{p=0}^{\ell}\bigoplus_{a=0}^{2-k}\bigoplus_{b=0}^{k}\;
				[p+a,\;\ell-p+b]_{\substack{-R_1(\ell-2p+3k+2a-2b-3)\\+R_2(n-\ell-2t-k-1)}}\; .	\nonumber\\
\end{eqnarray}
%

\section{Possible space invasion patterns} \label{app: scenarios}


Group theory alone is not enough to determine the precise structure of the full KK spectrum of GMPS, once the assumption that the arrangement into OSp$(4|2)$ supermultiplets should occur KK level by KK level is abandoned. Space invaders can be drawn from higher KK levels in multiple ways that are still compatible with group theory. In this appendix, we go through a couple of these possibilities. Short of computing the actual spectrum, the present analysis remains inconclusive about the precise invasion pattern that is realised in the spectrum. The possible invasion patterns discussed below have been determined using the group theory branchings of appendix \ref{sec: branchings}.

\subsection{Space invaders at level $n$ drawn from level $ n+1$} \label{sec:KKnandn+1} \label{sec:nAndn+1Pattern}

At KK level $n$, it typically happens that all the $\textrm{SU}(3) \times \textrm{U}(1)_3$ states at that level can be allocated into OSp$(4|2)$ multiplets, but these states are not enough to fill out these multiplets entirely. States in the same $\textrm{SU}(3)$ representation and with the appropriate $\textrm{U}(1)_3$ R-charges must be selected from higher KK levels in order to complete the multiplets. We think of the former states as naturally belonging to KK level $n$, while we refer to the latter states as space invaders. A working assumption consists in drawing invading states at KK level $n$ only from the immediately higher level, $n+1$, for all $n$. Using this prescription, it is possible to fill out OSp$(4|2)$ multiplets consistently, as we check in tables \ref{tab:multipletsatlevel0}, \ref{tab:multipletsatlevel1} and \ref{tab:multipletsatlevel2} for KK levels $n=0$, $n=1$ and $n=2$, respectively. Table \ref{table: branchKK0} in section \ref{sec: spaceinvaders}  contains further details of the case covered in table \ref{tab:multipletsatlevel0}. Likewise, table \ref{tab:FurtherDetB2} contains further details of the case covered in table \ref{tab:multipletsatlevel1}. We kindly borrow the format of tables \ref{table: branchKK0} through \ref{tab:DetailsFrommultipletsatlevel1from2and3} from \cite{Klebanov:2008vq}, as well as the OSp$(4|2)$ supermultiplet terminology. Also, in tables \ref{tab:multipletsatlevel0}, \ref{tab:multipletsatlevel1}, etc., for each OSp$(4|2)$ supermultiplet in the indicated SU(3) representations, the value of its R-charge is indicated.


\begin{table}[H]
\begin{center}
{\footnotesize
\begin{tabular}{|p{20mm}|p{20mm}|p{20mm}|} 					\hline
$[0,0]$ 				& 	$[0,1]$ 				& 	$[0,2]$ 			\\
MGRAV $0$ 			& SGINO$^\star$ $-\frac19$ 	& HYP $-\frac{8}{9}$ 	\\
LVEC$^\star$ $0$ 		& SVEC$^\star$ $+\frac29$	& 					\\
HYP $+\frac43,-\frac43$ 	& 						& 					\\ \hline
$[1,0]$ 				& 	$[1,1]$ 			\\
SGINO$^\star$ $+\frac19$& MVEC $0$ 			\\
SVEC$^\star$ $-\frac29$ 	& 		 			\\ \cline{1-2}
$[2,0]$ 			\\
HYP $+\frac{8}{9}$ 	\\ \cline{1-1}
\end{tabular}
}
\caption{Supermultiplets at KK level $n=0$.  A star denotes that the completion of the corresponding supermultiplet uses states coming from level $n=1$. See table \ref{table: branchKK0} in section \ref{sec: spaceinvaders} for further details.}
\label{tab:multipletsatlevel0}
\end{center}
\end{table}


\begin{table}[H]
\begin{center}
{\footnotesize
\begin{tabular}{|p{25mm}|p{25mm}|p{25mm}|p{25mm}|} \hline
$[0,0]$ 							& 	$[0,1]$ 				& 	$[0,2]$ 			& 	$[0,3]$ \\
SGRAV $+\frac23$, $-\frac23$			& LGRAV$^\star$ $-\frac49$ 	& SGINO $-\frac59$ 	& HYP $-\frac43$ \\
LVEC$^\star$ $+\frac23$, $-\frac23$		& LGINO$^\star$ $+\frac59$	& SVEC  $-\frac29$   	&  \\
                 						& SGINO $-\frac79$			& HYP $-\frac{14}9$		&  \\
                 						& HYP$^\star$ $-\frac{10}9$	& 					&  \\ \hline
$[1,0]$ 					& 	$[1,1]$ 					& 	$[1,2]$ 			\\
LGRAV$^\star$ $+\frac49$ 	& SGINO $+\frac13$, $-\frac13$	& SVEC $-\frac49$ 		\\
LGINO$^\star$ $-\frac59$		& SVEC $+\frac23$, $-\frac23$		&                    			\\
SGINO $+\frac79$			&           						&                    			\\
HYP$^\star$ $+\frac{10}9$	&           						&                    			\\ \cline{1-3}
$[2,0]$ 			& 	$[2,1]$ 				\\
SGINO $+\frac59$ 	& SVEC $+\frac49$	 		\\
SVEC $+\frac29$ 	& 				 		\\
HYP  $+\frac{14}9$ &                    \\ \cline{1-2}
$[3,0]$ \\
HYP $+\frac43$ \\ \cline{1-1}
\end{tabular}
}
\caption{Supermultiplets at KK level $n=1$.  A star denotes that the completion of the corresponding supermultiplet uses states coming from level $n=2$. See table \ref{tab:FurtherDetB2} for further details.}
\label{tab:multipletsatlevel1}
\end{center}
\end{table}


\begin{table}[H]
\begin{center}
{\footnotesize
\begin{tabular}{|p{32mm}|p{25mm}|p{25mm}|p{25mm}|p{25mm}|} \hline
$[0,0]$ 								& 	$[0,1]$ 			& 	$[0,2]$ 			& 	$[0,3]$			& 	$[0,4]$ 		\\
LGRAV 0								& 				 	& 					& 					&				\\
SGRAV $+\frac43$, $-\frac43$				& 	conj. to [1,0]	 	& 	conj. to [2,0]		& 	conj. to [3,0]		&	conj. to [4,0]	\\
LVEC $+\frac43^\star$, $-\frac43^\star$, 0, 0	& 			 		& 					& 					&				\\\hline
$[1,0]$ 										& 	$[1,1]$ 						& 	$[1,2]$ 			&	$[1,3]$			\\
LGRAV $+\frac{10}9^\star$, $-\frac29$				& 	LGRAV $0$					& 					&					\\
LGINO $+\frac19$, $+\frac19^\star$, $-\frac{11}9$		& 	LGINO $\pm1^\star$, $\pm\frac13$	&      conj. to [2,1]		&	conj. to [3,1]		\\
LVEC $+\frac49$ 								& 	SVEC $+\frac43$, $-\frac43$		&      					&					\\
SVEC $-\frac29^\star$							&	LVEC $0$, $0$					&      					&					\\
HYP $+\frac{10}9^\star$, $+\frac{4}9^\star$			&								&&\\\cline{1-4}
$[2,0]$ 					&	$[2,1]$ 						&	$[2,2]$		\\
LGRAV $+\frac89^\star$		&	LGINO $+\frac79^\star$, $+\frac19$ 	&	LVEC $0$\\
SGINO $+\frac{11}9$	 	&	SVEC $+\frac{10}9$		&   \\
LGINO $-\frac19$		 	&	LVEC $-\frac29^\star$	&   \\
LVEC $+\frac29$, $-\frac49$	&						&   \\
HYP $+\frac{20}9$, $+\frac{8}9$&						&   \\ \cline{1-3}
$[3,0]$ 				&	$[3,1]$			\\
SGINO $+1$	 		&	SVEC $+\frac89$	\\
LVEC $+\frac23^\star$ 	&					\\
HYP $+2$		 		&					\\ \cline{1-2}
$[4,0]$ 			\\
HYP $\frac{16}9$	\\ \cline{1-1}
\end{tabular}
}
\caption{Supermultiplets at KK level $n=2$.  A star denotes that the completion of the corresponding supermultiplet uses states coming from level $n=3$.}
\label{tab:multipletsatlevel2}
\end{center}
\end{table}


\begin{table}
\begin{center}
\resizebox{\textwidth}{!}{
\begin{tabular}{|c|c|c|c|c|c|c|c|c|c|c|c|c|c|c|c|} \hline
Spin & SO(8) & \multicolumn{14}{c|}{$\text{SU}(3)\times\text{U}(1)_{3}$} \\ \hline
%
%
$2$ 	&	$\bm{8}_v$ & $\bm{1}_{\frac23}$ & $\bm{3}_{\frac49}$ &&&&&&&&&&&&   \\ \hline
%
%
$\frac{3}{2}$ & $\bm{56}_c$ & $\bm{1}_{-\frac13}$ & $\bm{3}_{-\frac59}$ & $\bm{8}_{\frac13}$ & $\bm{6}_{\frac59}$ & $\bm{3}_{\frac79}$ & $\bm{3}_{-\frac59}$ && &&&&&&\\
&$+$& $\bm{1}_{-\frac13}$ & $\bm{3}_{-\frac59}$ &&&&&&&&&&&&  \\  
& $\bm{8}_c$ & $\bm{1}_{\frac53}$ & $\bm{3}_{\frac{13}9}$ &&&&&&&&&&&&  \\ \cline{2-16}
& \Rocket & $$ & $\bm{3}_{\frac{13}9}$ &&&&&&&&&&&&  \\\hline
%
%
$1$ & $\bm{160}_v$ & $\bm{1}_{-\frac43}$ & $\bm{3}_{-\frac{14}9}$ & $\bm{8}_{-\frac23}$ & $\bm{6}_{-\frac49}$ & $\bm{3}_{-\frac29}$ & $\bm{3}_{-\frac{14}9}$
	& $\bm{15}_{\frac49}$ & $\bm{8}_{\frac23}$ & $\bm{6}_{\frac29}$ & $\bm{1}_{\frac23}$ & &&&$\bm{3}_{-\frac89}$ \\
&$+$ & $\bm{1}_{\frac23}$ & $\bm{3}_{\frac49}$, $\bm{3}_{\frac49}$ & $\bm{8}_{-\frac23}$ & $\bm{6}_{-\frac49}$ & $\bm{3}_{-\frac29}$ & $\bm{3}_{\frac49}$ &&&&&&&& \\
&$\bm{56}_v$ & $\bm{1}_{\frac23}$ & $\bm{3}_{\frac49}$, $\bm{3}_{\frac49}$ & $\bm{8}_{\frac43}$ & $\bm{6}_{\frac{14}9}$ & $\bm{3}_{\frac{16}9}$ &&&&&&&&& \\\cline{2-16}
&\Rocket&& $\bm{3}_{\frac{22}9}$ & $$  && & $\bm{3}_{-\frac{14}9}$ &&&&&&&& \\
&&&&&&& $\bm{3}_{\frac{4}9}$ &&&&&&&& \\ \hline
%
%
$\frac{1}{2}$ & $\bm{224}_{vc}$ & $\bm{1}_{-\frac13}$ & $\bm{3}_{-\frac59}$, $\bm{3}_{-\frac59}$ & $\bm{8}_{-\frac53}$ & $\bm{6}_{-\frac{13}9}$ 
	& $\bm{3}_{-\frac{11}9}$ & $\bm{3}_{-\frac59}$ & $\bm{15}_{-\frac59}$ & $\bm{8}_{-\frac13}$ & $\bm{6}_{-\frac79}$ & $\bm{1}_{-\frac13}$ & $\bm{10}_{\frac13}$ & $\bm{6}_{\frac{5}9}$   
	& $\bm{3}_{\frac19}$ &$\bm{3}_{-\frac{17}9}$, $\bm{3}_{\frac19}$, $\bm{3}_{\frac79}$ \\
&$+$&& $\bm{3}_{\frac{13}9}$, $\bm{3}_{\frac{13}9}$ & $\bm{8}_{\frac13}$ & $\bm{6}_{\frac59}$ & $\bm{3}_{\frac79}$ & $\bm{3}_{-\frac59}$ & $\bm{15}_{-\frac{5}9}$ & $\bm{8}_{-\frac13}$ 
	& $\bm{6}_{-\frac79}$ & $\bm{1}_{-\frac13}$ &&&& $\bm{1}_{-1}$, $\bm{1}_{1}$ \\
&$\bm{160}_{c}$&& $$ & $\bm{8}_{\frac13}$ & $\bm{6}_{\frac59}$ & $\bm{3}_{\frac79}$ & $\bm{3}_{-\frac59}$  & $\bm{15}_{\frac{13}9}$ & $\bm{8}_{\frac53}$ & $\bm{6}_{\frac{11}9}$ 
	& $\bm{1}_{\frac53}$ &&&&  \\\cline{2-16}
&\Rocket& & $$  & $$ && $$ & $\bm{3}_{-\frac{23}9}$ &&&& $\bm{1}_{\frac53}$ &&&$$&	\\
&&&&&&& $\bm{3}_{\frac{13}9}$ & $$ &&&&&$$&& \\ \hline

%
%
$0$ & $\bm{112}_v$ &  $$ & $\bm{3}_{\frac49}$ & $\bm{8}_{-\frac23}$ & $$& $$ & $$ & $\bm{15}_{\frac49}$ & $$ & $\bm{6}_{\frac29}$ & $\bm{1}_{\frac23}$ & $\bm{10}_{\frac43}$ 
	& $\bm{6}_{\frac{14}9}$ &  & $\bm{3}_{-\frac89}$, $\bm{3}_{\frac{16}9}$ \\
&&& $$ & $$ &&&&&&&&& $$ && $\bm{1}_{-2}$, $\bm{1}_{2}$ \\\cline{2-16}
%
%
& $\bm{224}_{cv}$ &  $$ & $$ & $$& $\bm{6}_{-\frac49}$ & $\bm{3}_{-\frac29}$ & $\bm{3}_{-\frac{14}9}$ & $\bm{15}_{\frac49}$ & $\bm{8}_{-\frac43}$ & $\bm{6}_{-\frac{16}9}$ 
	& $\bm{1}_{-\frac43}$ & $\bm{10}_{-\frac23}$ & $\bm{6}_{-\frac49}$ & $\bm{3}_{\frac{10}9}$ & $\bm{3}_{-\frac89}$, $\bm{3}_{-\frac29}$\\
&&& $$ & $$ &&& $\bm{3}_{\frac49}$ & $\bm{15}_{-\frac{14}9}$ & $\bm{8}_{\frac23}$ & $\bm{6}_{\frac29}$ & $\bm{1}_{\frac23}$ &  &&& $\bm{1}_{0}$, $\bm{1}_{0}$ \\
&&&&&&& $\bm{3}_{\frac49}$ & $$ & $\bm{8}_{\frac23}$ &&&  &  $$ &&\\\cline{2-16}
&\Rocket&& $$ & $$  && $$ & $\bm{3}_{-\frac{14}9}$ && $$ && $\bm{1}_{\frac23}$ &&& $\bm{3}_{-\frac89}$ &\\ 
&&&&&&&& $$ &&&$\bm{1}_{\frac83}$&&  $$ &&\\ \hline

\multicolumn{2}{l|}{} &
\rotatebox{90}{\mbox{Short graviton\;}} &
\rotatebox{90}{\mbox{Long graviton\;}} &
\rotatebox{90}{\mbox{Short gravitino\;}} &
\rotatebox{90}{\mbox{Short gravitino\;}} &
\rotatebox{90}{\mbox{Short gravitino\;}} &
\rotatebox{90}{\mbox{Long gravitino\;}} &
\rotatebox{90}{\mbox{Short vector\;}} &
\rotatebox{90}{\mbox{Short vector\;}} &
\rotatebox{90}{\mbox{Short vector\;}} &
\rotatebox{90}{\mbox{Long vector\;}} &
\rotatebox{90}{\mbox{Massive hyper\;}} &
\rotatebox{90}{\mbox{Massive hyper\;}} &
\rotatebox{90}{\mbox{Massive hyper\;}} &
\rotatebox{90}{\mbox{Invaders for $n=0$\;}}  \\  \cline{3-16}
\end{tabular}
}
\caption{Details of the branching of the $\mathcal{N}=8$ supermultiplets at KK level $n=1$ into Osp$(4\vert2)$ multiplets in SU(3)$\times$U(1)${}_3$ representations, as given in table \ref{tab:multipletsatlevel1}. \protect\Rocket\ denotes states coming from KK level $n=2$. The last column shows the states which were already needed to complete supermultiplets at KK level $n=0$. For every complex representation, the presence of its conjugate is understood.}
\label{tab:FurtherDetB2}
\end{center}
\end{table}

\newpage

\subsection{Matching the CPW short multiplets} \label{sec:MatchingCPW}


The invasion pattern proposed in section \ref{sec:KKnandn+1} leads to a number of short multiplets that do not have a counterpart for CPW, as can be checked by comparing tables \ref{tab:multipletsatlevel0}, \ref{tab:multipletsatlevel1} and \ref{tab:multipletsatlevel2} above with tables 17, 18 and 19 of \cite{Klebanov:2008vq}. We can turn these possible short multiplets in the GMPS spectrum into long ones ({\it i.e.~elongate} them) by putting in further extra states. At KK level $n$, these must necessarily involve KK levels higher than $n+1$. With the additional prescription that, at level $n$, we use as many invaders from level $n+1$ as possible, we find that these elongations are consistently possible by retrieving invaders from level $n+2$ only, with no other levels needed. Tables \ref{tab:multipletsatlevel0from1and2}, \ref{tab:multipletsatlevel1from2and3} and \ref{tab:multipletsatlevel2from3and4} illustrate this invasion pattern for KK levels $n=0$, $n=1$ and $n=2$, respectively. Further details on the allocation of supermultiplets of tables \ref{tab:multipletsatlevel0from1and2} and \ref{tab:multipletsatlevel1from2and3} can be found in tables \ref{tab:DetailsFrommultipletsatlevel0from1and2} and \ref{tab:DetailsFrommultipletsatlevel1from2and3}, respectively. Some ambiguities that arise using this prescription are discussed in the caption of table \ref{tab:multipletsatlevel2from3and4}.


\begin{table}[H]
\begin{center}
{\footnotesize
\begin{tabular}{|p{22mm}|p{20mm}|p{20mm}|} 					\hline
$[0,0]$ 								& 	$[0,1]$ 					& 	$[0,2]$ 			\\
MGRAV $0$ 							& SGINO $-\frac19^{\star}$ 		& HYP $-\frac{8}{9}$ 	\\
LVEC $0^{\star}$, $\pm\frac23^{\star\star}$ 	& LVEC $+\frac29^{\star\star}$			& 					\\ \hline
$[1,0]$ 					& 	$[1,1]$ 			\\
SGINO $+\frac19^{\star}$		& 	MVEC $0$ 		\\
LVEC $-\frac29^{\star\star}$ 	& 		 			\\ \cline{1-2}
$[2,0]$ 			\\
HYP $+\frac{8}{9}$ 	\\ \cline{1-1}
\end{tabular}
}
\caption{Multiplets at level $n=0$ with short multiplets in the same SU(3) representations as \cite{Klebanov:2008vq}. A star now denotes that the multiplet involves states coming from level $n=1$, with a second star denoting that states from level $n=2$ are also used. See table \ref{tab:DetailsFrommultipletsatlevel0from1and2} for further details.}
\label{tab:multipletsatlevel0from1and2}
\end{center}
\end{table}


\begin{table}[H]
\begin{center}
{\footnotesize
\begin{tabular}{|p{30mm}|p{30mm}|p{25mm}|p{25mm}|} \hline
$[0,0]$ 							& 	$[0,1]$ 								& 	$[0,2]$ 								& 	$[0,3]$ \\
SGRAV $\pm\frac23$				& LGRAV $-\frac49^\star$ 					& SGINO $-\frac59$ 							& HYP $-\frac43$ \\
LVEC $\pm\frac43^{\star\star}$			& LGINO $+\frac59^\star$, $-\frac79^{\star\star}$	& LVEC  $-\frac29^\star$, $\frac{4}9^{\star}$  	&  \\
                 						& LVEC $-\frac{10}9^{\star\star}$				&  		&  \\ \hline
$[1,0]$ 									& 	$[1,1]$ 					& 	$[1,2]$ 			\\
LGRAV $+\frac49^\star$ 						& LGINO $\pm \frac13^{\star}$	& SVEC $-\frac49$ 		\\
LGINO $-\frac59^\star$, $+\frac79^{\star\star}$		& LVEC $\pm\frac23^{\star\star}$		&                    			\\
LVEC $+\frac{10}9^{\star\star}$					&           						&                    			\\ \cline{1-3}
$[2,0]$ 									& 	$[2,1]$ 				\\
SGINO $+\frac59$ 							& SVEC $+\frac49$	 		\\
LVEC $+\frac29^\star$, $-\frac{4}9^{\star}$		& 				 		\\ \cline{1-2}
$[3,0]$ \\
HYP $+\frac43$ \\ \cline{1-1}
\end{tabular}
}
\caption{Multiplets at level $n=1$ with short multiplets in the same SU(3) representations as \cite{Klebanov:2008vq}. A star now denotes that the multiplet involves states coming from level $n=2$, with a second star denoting that states from level $n=3$ are also used. See table \ref{tab:DetailsFrommultipletsatlevel1from2and3} for further details.}
\label{tab:multipletsatlevel1from2and3}
\end{center}
\end{table}


\begin{table}[H]
\begin{center}
{\footnotesize
\begin{tabular}{|p{36mm}|p{28mm}|p{25mm}|p{25mm}|p{25mm}|} \hline
$[0,0]$ 							& 	$[0,1]$ 			& 	$[0,2]$ 			& 	$[0,3]$			& 	$[0,4]$ 		\\
LGRAV 0							& 				 	& 					& 					&				 \\
SGRAV $+\frac43$, $-\frac43$			& 	conj. to [1,0]	 	& 	conj. to [2,0]		& 	conj. to [3,0]		&	conj. to [4,0]	 \\
LVEC  0, 0, ...						& 				 		& 				& 					&				 \\\hline
$[1,0]$ 											& 	$[1,1]$ 								& 	$[1,2]$ 		&	$[1,3]$			\\
LGRAV $+\frac{10}9^\star$, $-\frac29^\star$					& 	LGRAV $0$							& 				&					\\
LGINO $+\frac19$, $+\frac19^\star$, $-\frac{11}9^\star$	& 	LGINO $\pm1^\star$, $\pm\frac13^\star$		&      conj. to [2,1]	&	conj. to [3,1]		\\
LVEC $+\frac49^\star$, $-\frac29^{\star\star}$					& 	LVEC $0^\star$, $0^\star$					&				&	\\\cline{1-4}
$[2,0]$ 									&	$[2,1]$ 									&	$[2,2]$		\\
LGRAV $+\frac89^\star$						&	LGINO $+\frac79^\star$, $+\frac19$ 				&	LVEC $0$		\\
LGINO $+\frac{11}9^{\star\star}$, $-\frac19^\star$	&	LVEC $+\frac{10}9^{\star\star}$, $-\frac29^{\star}$	&				\\
LVEC $+\frac29^{\star\star}$, $-\frac49^\star$	, $\frac89^\star$	&											&				\\ \cline{1-3}
$[3,0]$ 							&	$[3,1]$			\\
SGINO $+1$	 					&	SVEC $+\frac89$	\\
LVEC $+\frac23^\star$, $0^{\star\star}$ 	&					\\ \cline{1-2}
$[4,0]$ 			\\
HYP $+\frac{16}9$	\\ \cline{1-1}
\end{tabular}
}
\caption{Multiplets at level $n=2$ with short multiplets in the same SU(3) representations as \cite{Klebanov:2008vq}. A star now denotes that the multiplet involves states coming from level $n=3$, with a second star denoting that states from level $n=4$ are also used. There is an ambiguity for the $[1,0]$ states: an invader from level $n=4$ could either complete a LVEC $-\frac29$ or a LGINO $-\frac{11}{9}$: we arbitrarily opted for the first choice. Furthermore, the dots in the SU(3) singlets denote that there is a leftover state that could belong to any long multiplet.}
\label{tab:multipletsatlevel2from3and4}
\end{center}
\end{table}



\begin{table}
\begin{center}
\resizebox{1.05\textwidth}{!}{
\begin{tabular}{|c|c|c|c|c|c|c|c|c|c|c|c|c|c|} \hline
Spin & SO(8) & \multicolumn{12}{c|}{$\text{SU}(3)\times\text{U}(1)_{3}$} \\ \hline
%
%
$2$ 	&	$\bf{1}$ & 		$\bm{1}_{0}$ &&&&&&&&&&&   \\ \hline
%
%
$\frac{3}{2}$	&	$\bm{8}_s$	& $\bm{1}_{+1}$ 	& $\bm{3}_{\frac19}$ & $\bm{\bar{3}}_{-\frac19}$ &  &  &&& &&&&\\
&&$\bm{1}_{-1}$ 	&&&&&&&&&&&  \\  \hline
%
%
$1$ & $\bm{28}$ &  $\bm{1}_{0}$ & $\bm{3}_{-\frac89}$ & $\bm{\bar{3}}_{\frac89}$& $\bm{8}_{0}$ &$\bm{3}_{-\frac29}$&$\bm{\bar{3}}_{\frac29}$&$\bm{1}_0$&&&&& \\
&&& $\bm{3}_{\frac{10}9}$ & $\bm{\bar{3}}_{-\frac{10}9}$ &&&&&&&&& \\\cline{2-14}
&\Rocket$_{\,1}$&& $\bm{3}_{-\frac89}$ & $\bm{\bar{3}}_{\frac89}$  &&&&&$\bm{1}_{-\frac23}$&$\bm{1}_{\frac23}$&&& \\ \hline
%
%
$\frac{1}{2}$ & $\bm{56}_s$ &   & $\bm{3}_{\frac19}$ & $\bm{\bar{3}}_{-\frac19}$ & $\bm{8}_{+1}$ &$\bm{3}_{\frac79}$& $\bm{\bar{3}}_{-\frac79}$ & 
	$\bm{1}_{+1}$ &$\bm{1}_{\frac13}$&$\bm{1}_{-\frac13}$&$\bm{6}_{-\frac19}$&$\bm{\bar{6}}_{\frac19}$&  \\
&&&  $\bm{3}_{\frac19}$ & $\bm{\bar{3}}_{-\frac19}$ & $\bm{8}_{-1}$ & $\bm{3}_{-\frac{11}9}$ & $\bm{\bar{3}}_{\frac{11}9}$  & $\bm{1}_{-1}$ &&&&& \\\cline{2-14}
&\Rocket$_{\,1}$& & $\bm{3}_{-\frac{17}9}$  & $\bm{\bar{3}}_{\frac{17}9}$ && $\bm{3}_{\frac79}$ & $\bm{\bar{3}}_{-\frac79}$ 
	& $\bm{1}_{+1}$ &$\bm{1}_{\frac13}$&$\bm{1}_{-\frac13}$&&&$\bm{3}_{\frac19}$	\\
&&&&&&$\bm{3}_{-\frac{11}9}$&$\bm{\bar{3}}_{\frac{11}9}$& $\bm{1}_{-1}$ &$\bm{1}_{-\frac53}$&$\bm{1}_{\frac53}$&&&$\bm{\bar{3}}_{-\frac19}$ \\\cline{2-14}
&\Rocket$_{\,2}$&&&&&&&  &$\bm{1}_{-\frac53}$&$\bm{1}_{\frac53}$&&& \\ \hline

%
%
$0$ & $\bm{35}_v$ &  $$ & $$ & $$& $\bm{8}_{0}$ & $\bm{3}_{-\frac29}$ & $\bm{\bar{3}}_{\frac29}$ & $\bm{1}_{0}$ &$\bm{1}_{\frac43}$	&$\bm{1}_{-\frac43}$
	&$\bm{6}_{\frac89}$&$\bm{\bar{6}}_{-\frac89}$& $\bm{3}_{\frac{10}9}$ \\
&&& $$ & $$ &&&&&&&&& $\bm{\bar{3}}_{-\frac{10}9}$\\\cline{2-14}
%
%
& $\bm{35}_c$ &  $$ & $\bm{3}_{-\frac89}$ & $\bm{\bar{3}}_{\frac89}$& $\bm{8}_{0}$ & $\bm{3}_{-\frac29}$ & $\bm{\bar{3}}_{\frac29}$ & $\bm{1}_{0}$ 
	&$\bm{1}_{-\frac23}$&$\bm{1}_{\frac23}$&$\bm{6}_{-\frac{10}9}$&$\bm{\bar{6}}_{\frac{10}9}$& \\
&&& $$ & $$ &&&&&&&&& \\\cline{2-14}
&\Rocket$_{\,1}$&& $$ & $$  && $\bm{3}_{\frac{16}9}$ & $\bm{\bar{3}}_{-\frac{16}9}$ & $\bm{1}_{0}$ &$\bm{1}_{-\frac23}$&$\bm{1}_{\frac23}$&&& $\bm{3}_{-\frac89}$, $\bm{\bar{3}}_{\frac89}$ \\ 
&&&&&&$\bm{3}_{-\frac29}$&$\bm{\bar{3}}_{\frac29}$& $\bm{1}_{+2}$ &$\bm{1}_{-\frac23}$&$\bm{1}_{\frac23}$&&&  $\bm{3}_{-\frac89}$, $\bm{\bar{3}}_{\frac89}$ \\
&&&&&&&& $\bm{1}_{-2}$ &&&&&  $\bm{3}_{-\frac29}$, $\bm{\bar{3}}_{\frac29}$\\ 
&&&&&&&&&&&&& $\bm{1}_0$ \\ \cline{2-14}
&\Rocket$_{\,2}$&&&&&$\bm{3}_{-\frac{20}9}$&$\bm{\bar{3}}_{\frac{20}9}$&&$\bm{1}_{-\frac83}$&$\bm{1}_{\frac83}$&&&  $\bm{1}_{-\frac23}$, $\bm{1}_{\frac23}$\\ \hline
\multicolumn{2}{l|}{} &
\rotatebox{90}{\mbox{Massless graviton\;}} &
\rotatebox{90}{\mbox{Short gravitino\;}} &
\rotatebox{90}{\mbox{Short gravitino\;}} &
\rotatebox{90}{\mbox{Massless vector\;}} &
\rotatebox{90}{\mbox{Long vector\;}} &
\rotatebox{90}{\mbox{Long vector\;}} &
\rotatebox{90}{\mbox{Long vector\;}} &
\rotatebox{90}{\mbox{Long vector\;}} &
\rotatebox{90}{\mbox{Long vector\;}} &
\rotatebox{90}{\mbox{Massive hyper\;}} &
\rotatebox{90}{\mbox{Massive hyper\;}} &
\rotatebox{90}{\mbox{Eaten modes\;}} \\  \cline{3-14}
\end{tabular}
}
\caption{Branching of the $\mathcal{N}=8$ massless graviton multiplet into Osp$(4\vert2)$ multiplets in SU(3)$\times$U(1)${}_3$ representations with short multiplets in the same SU(3) representations as in \cite{Klebanov:2008vq}, as summarised in table \ref{tab:multipletsatlevel0from1and2}. \protect\Rocket$_{\,n}$\  denotes states coming from KK levels $n=1,2$.}
\label{tab:DetailsFrommultipletsatlevel0from1and2}
\end{center}
\end{table}

\begin{table}
\begin{center}
\resizebox{1.05\textwidth}{!}{
\begin{tabular}{|c|c|c|c|c|c|c|c|c|c|c|c|c|c|c|c|} \hline
Spin & SO(8) & \multicolumn{14}{c|}{$\text{SU}(3)\times\text{U}(1)_{3}$} \\ \hline
%
%
$2$ 	&	$\bf{8}_v$ & 		$\bm{1}_{\frac 23}$ & $\bm{3}_{\frac 49}$&&&&&&&&&&&&\\ \hline
%
%
$\frac{3}{2}$	&	$\bm{56}_c$ & $\bm{1}_{\frac 53}$ & $\bm{3}_{-\frac 59}$ & $\bm{6}_{\frac 59}$ & $\bm{8}_{\frac 13}$ & $\bm{3}_{-\frac 59}$ & $\bm{3}_{\frac 79}$&&&&&&&& \\
                              &+& $\bm{1}_{-\frac 13}$ &  $\bm{3}_{\frac {13}{9}}$&&&&&&&&&&&&\\    
                                                           &$\bm{8}_c$& $\bm{1}_{-\frac 13}$ & $\bm{3}_{-\frac 59}$&&&&&&&&&&&&\\ \cline{2-16} 
&\Rocket$_{\,2}$&&$\bm{3}_{\frac {13}{9}}$&&&&&&&&&&&&\\ \hline   
%
$1$	&	& $\bm{1}_{\frac 23}$ &  $\bm{3}_{\frac 49}$ &  $\bm{6}_{\frac {14}{9}}$ &  $\bm{8}_{\frac 43}$ & $\bm{3}_{\frac 49}$&  $\bm{3}_{-\frac 29}$ &  $\bm{15}_{\frac 49}$ &&$\bm{8}_{\frac 23}$&  $\bm{6}_{\frac 29}$&&&& $\bm{3}_{-\frac 89}$ \\
                                     &$\bm{160}_v$ & $\bm{1}_{-\frac 43}$& $\bm{3}_{\frac 49}$ &  $\bm{6}_{-\frac 49}$ &  $\bm{8}_{-\frac 23}$&$\bm{3}_{-\frac {14}{9}}$&  $\bm{3}_{\frac {16}{9}}$&&&&&&&& $\bm{1}_{\frac 23}$\\
 &+&$\bm{1}_{\frac 23}$& $\bm{3}_{-\frac {14}{9}}$& $\bm{6}_{-\frac 49}$& $\bm{8}_{-\frac 23}$&& $\bm{3}_{-\frac 29}$&&&&&&&&$\bm{1}_{-\frac 23}$     \\                                  
                                    &$\bm{56}_v$&&$\bm{3}_{\frac 49}$& & & &&&&&&&&&\\ 
 && & $\bm{3}_{\frac 49}$&&&&&&&&&&&&     \\  \cline{2-16} 
&\Rocket$_{\,2}$&&$\bm{3}_{\frac{22}{9}}$&& $\bm{8}_{\frac 43}$& $\bm{3}_{\frac 49}$& $\bm{3}_{\frac {16}{9}}$&& $\bm{1}_{-\frac 43}$&&& $\bm{6}_{-\frac 49}$& $\bm{3}_{\frac {10}{9}}$&&\\ 
&&&&&& $\bm{3}_{-\frac {14}{9}}$&&&&&&&&&\\ \hline  
%
$\frac{1}{2}$	&		& $\bm{1}_{-\frac 13}$&$\bm{3}_{\frac{13}{9}}$&$\bm{6}_{\frac 59}$&$\bm{8}_{\frac 13}$&$\bm{3}_{-\frac 59}$&$\bm{3}_{\frac 79}$&$\bm{15}_{\frac {13}{9}}$&$\bm{1}_{-\frac 13}$&$\bm{8}_{\frac 53}$&$\bm{6}_{-\frac 79}$&$\bm{6}_{\frac 59}$&$\bm{3}_{\frac 19}$&$\bm{10}_{\frac 13}$&$\bm{3}_{-\frac{17}{9}}$, $\bm{3}_{-\frac{11}{9}}$\\ 
                                                     &$\bm{224}_{vc}$&&$\bm{3}_{-\frac{5}{9}}$& $\bm{6}_{\frac 59}$& $\bm{8}_{\frac 13}$&$\bm{3}_{-\frac 59}$&$\bm{3}_{\frac 79}$& $\bm{15}_{-\frac 59}$&&&$\bm{6}_{\frac {11}{9}}$&&&&$\bm{3}_{\frac 79}$ , $\bm{3}_{\frac 19}$  \\ 
                                                         &+&&$\bm{3}_{-\frac{5}{9}}$&$\bm{6}_{-\frac {13}{9}}$&$\bm{8}_{-\frac 53}$& $\bm{3}_{-\frac 59}$&&$\bm{15}_{-\frac 59}$&&&$\bm{6}_{-\frac 79}$&&&&$\bm{1}_{-1}$, $\bm{1}_{\frac 53}$ \\ 
                                                         &$\bm{160}_{c}$&&$\bm{3}_{\frac {13}{9}}$&&$\bm{8}_{\frac 13}$&&&&&&&&&& $\bm{1}_{1}$, $\bm{1}_{-\frac 53}$ \\
                                                         &&&&&$\bm{8}_{\frac 13}$&&&&&&&&&&$\bm{1}_{-\frac 13}$, $\bm{1}_{\frac 13}$ \\    \cline{2-16}                                                
                                                       
&\Rocket$_{\,2}$&&&& $\bm{8}_{\frac 73}$&$\bm{3}_{-\frac 59}$& $\bm{3}_{\frac 79}$&& $\bm{1}_{-\frac 13}$& $\bm{8}_{-\frac 13}$& $\bm{6}_{\frac{11}{9}}$& $\bm{6}_{\frac 59}$& $\bm{3}_{\frac 19}$&&\\
                                                    &&&&&&$\bm{3}_{\frac{13}{9}}$& $\bm{3}_{\frac{7}{9}}$&&$\bm{1}_{-\frac 73}$&$\bm{8}_{-\frac 13}$&& $\bm{6}_{-\frac{13}{9}}$&$\bm{3}_{\frac{19}{9}}$&&\\
                                                    &&&&&&$\bm{3}_{-\frac{23}{9}}$& $\bm{3}_{-\frac{11}{9}}$&&&$\bm{8}_{\frac 53}$&& $\bm{6}_{-\frac{13}{9}}$&$\bm{3}_{\frac{19}{9}}$&&\\ \cline{2-16} 
&\Rocket$_{\,3}$&&&&&& $\bm{3}_{\frac{25}{9}}$&& $\bm{1}_{-\frac 73}$&&&&&&\\ \hline  
%
$0$	&	$\bm{112}_v$&& $\bm{3}_{\frac 49}$&& $\bm{8}_{-\frac 23}$&&& $\bm{15}_{\frac 49}$&&& $\bm{6}_{\frac 29}$& $\bm{6}_{\frac{14}{9}}$&& $\bm{10}_{\frac 43}$& $\bm{3}_{\frac{16}{9}}$, $\bm{3}_{-\frac 89}$ \\ 
                                     &&&&&&&&&&&&&&& $\bm{1}_{\frac 23}$, $\bm{1}_{2}$\\ 
                                     &&&&&&&&&&&&&&& $\bm{1}_{-\frac 23}$, $\bm{1}_{-2}$\\ \cline{2-16}

	&	$\bm{224}_{cv}$&&&$\bm{6}_{-\frac 49}$& $\bm{8}_{-\frac 23}$& $\bm{3}_{\frac 49}$&& $\bm{15}_{\frac 49}$&$\bm{1}_{-\frac 43}$&  $\bm{8}_{\frac 23}$& $\bm{6}_{\frac 29}$& $\bm{6}_{-\frac 49}$& $\bm{3}_{\frac{10}{9}}$& $\bm{10}_{-\frac 23}$&  $\bm{3}_{-\frac 89}$ \\ 
                                     &&&&& $\bm{8}_{\frac 43}$& $\bm{3}_{\frac 49}$&& $\bm{15}_{-\frac {14}{9}}$&&& $\bm{6}_{-\frac {16}{9}}$&&&&$\bm{3}_{-\frac 29}$, $\bm{3}_{-\frac 29}$  \\
                                    &&&&&& $\bm{3}_{-\frac{14}{9}}$&&&&&&&&&$\bm{1}_{0}$, $\bm{1}_{0}$\\ 
                                    &&&&&&&&&&&&&&&$\bm{1}_{\frac 23}$, $\bm{1}_{-\frac 23}$\\    \cline{2-16}
&\Rocket$_{\,2}$&&&& $\bm{8}_{\frac 43}$& $\bm{3}_{-\frac{14}{9}}$ &$\bm{3}_{\frac{16}{9}}$&&$\bm{1}_{\frac 23}$& $\bm{8}_{-\frac 43}$& $\bm{6}_{\frac 29}$ &$\bm{6}_{-\frac{22}{9}}$& $\bm{3}_{-\frac 89}$ &&\\ 
                                   &&&&&&& $\bm{3}_{-\frac 29}$&& $\bm{1}_{-\frac 43}$&$\bm{8}_{\frac 23}$&$\bm{6}_{\frac{20}{9}}$& $\bm{6}_{-\frac 49}$& $\bm{3}_{\frac {10}{9}}$&&\\
                                   &&&&&&& $\bm{3}_{-\frac 29}$&& $\bm{1}_{-\frac 43}$&$\bm{8}_{\frac 23}$&& $\bm{6}_{-\frac 49}$& $\bm{3}_{\frac {10}{9}}$&&\\ \cline{2-16} 
&\Rocket$_{\,3}$&&&&&& $\bm{3}_{\frac{16}{9}}$&& $\bm{1}_{-\frac {10}{3}}$&$\bm{8}_{\frac 83}$&&& $\bm{3}_{\frac{28}{9}}$&&\\ \hline 
\multicolumn{2}{l|}{} &
\rotatebox{90}{\mbox{Short graviton\;}} &
\rotatebox{90}{\mbox{Long graviton\;}} &
\rotatebox{90}{\mbox{Short gravitino\;}} &
\rotatebox{90}{\mbox{Long gravitino\;}} &
\rotatebox{90}{\mbox{Long gravitino\;}} &
\rotatebox{90}{\mbox{Long gravitino\;}} &
\rotatebox{90}{\mbox{Short vector\;}} &
\rotatebox{90}{\mbox{Long vector\;}} &
\rotatebox{90}{\mbox{Long vector\;}} &
\rotatebox{90}{\mbox{Long vector\;}} &
\rotatebox{90}{\mbox{Long vector\;}} &
\rotatebox{90}{\mbox{Long vector\;}} &
\rotatebox{90}{\mbox{Massive hyper\;}} &
\rotatebox{90}{\mbox{Invaders for n=0\;}} \\  \cline{3-16}

\end{tabular}
}
\caption{Branching of the $\mathcal{N}=8$ supermultiplets at KK level $n=1$ into Osp$(4\vert2)$ multiplets in SU(3)$\times$U(1)${}_3$ representations, as summarised in table \ref{tab:multipletsatlevel1from2and3}. \protect\Rocket$_{\,n}$ denotes states coming from KK level $n=2,3$. The last column shows the states which were already needed to complete supermultiplets at KK level $n=0$. For every complex representation, the presence of its conjugate is understood.}
\label{tab:DetailsFrommultipletsatlevel1from2and3}
\end{center}
\end{table}

\newpage

\bibliography{references}
\end{document}